\documentclass{llncs}
\usepackage{graphicx}

\usepackage{amssymb,amsmath}
\usepackage{upgreek}
\usepackage[all]{xy}
\usepackage{stmaryrd}
\usepackage{code}
\usepackage{multirow}
\usepackage{caption}

\newcommand{\defterm}[1]{\textbf{#1}}

\newcommand{\ie}{\emph{i.e.}}
\newcommand{\eg}{\emph{e.g.}}

\newcommand{\etal}{\emph{et al.}}

\newcommand{\syn}[1]{\mathsf{#1}}
\newcommand{\var}[1]{\mathit{#1}}
\newcommand{\s}[1]{\mathit{#1}}

\newcommand{\parto}{\rightharpoonup}

\newcommand{\set}[1]{\left\{#1\right\}}
\newcommand{\setbuild}[2]{\left\{ #1 : #2\right\}}

\newcommand{\Pow}[1]{{\mathcal{P}\left(#1\right)}}
\newcommand{\PowSm}[1]{{\mathcal{P}(#1)}}

\newcommand{\union}{\cup}

\newcommand{\vect}[1]{\langle #1\rangle}
\newcommand{\vecp}[1]{\vec{#1}\;'}

\newcommand{\To}{\mathrel{\Rightarrow}}

\newcommand{\join}{\sqcup}

\newcommand{\bigjoin}{\bigsqcup}

\newcommand{\QStates}{Q}
\newcommand{\FStates}{F}

\newcommand{\StackAlpha}{\Gamma}
\newcommand{\stackchar}{\gamma}

\newcommand{\sembr}[1]{\ensuremath{[\![{#1}]\!]}}

\newcommand{\opor}{\mathrel{|}}

\newcommand{\Alphabet}{A}

\newcommand{\produces}{\mathrel{::=}}

\newcommand{\stmt}{s}

\newcommand{\Stmt}{\syn{Stmt}}

\newcommand{\expr}{e}
\newcommand{\aexpr}{\mbox{\sl {\ae}}}

\newcommand{\op}{{\var{op}}}

\newcommand{\Eval}{{\mathcal{E}}}
\newcommand{\ArgEval}{{\mathcal{A}}}

\newcommand{\Inject}{{\mathcal{I}}}

\newcommand{\qstate}{q}

\newcommand{\Store}{\s{Store}}

\newcommand{\den}{d}

\newcommand{\store}{\sigma}

\newcommand{\cont}{\kappa}

\newcommand{\addr}{a}

\newcommand{\Lives}{\mathit{Lives}}

\newcommand{\aTo}{\leadsto}

\newcommand{\aInject}{{\hat{\mathcal{I}}}}

\newcommand{\sa}[1]{\widehat{\mathit{#1}}}

\newcommand{\aEval}{{\hat{\mathcal{E}}}}
\newcommand{\aArgEval}{{\hat{\mathcal{A}}}}

\newcommand{\aStore}{\sa{Store}}

\newcommand{\astore}{{\hat{\sigma}}}

\newcommand{\acont}{{\hat{\kappa}}}

\newcommand{\aden}{{\hat{d}}}

\newcommand{\aaddr}{{\hat{\addr}}}

\newcommand{\cexpr}{\mathit{ce}}

\newcommand{\afp}{\hat{\mathit{fp}}}

\newcommand{\reg}{\mathit{name}}

\newcommand{\ControlStates}{Q}
\newcommand{\transfunction}{\delta}

\newcommand{\conf}{c}
\newcommand{\aconf}{{\hat c}}

\newcommand{\phrame}{\phi}
\newcommand{\aphrame}{\hat{\phi}}

\newcommand{\callframe}{\chi}
\newcommand{\acallframe}{\widehat{\chi}}
\newcommand{\exnframe}{\eta}
\newcommand{\aexnframe}{\widehat{\eta}}

\newcommand{\stackact}{g}

\DeclareMathOperator*{\PDTrans}{\longmapsto}

\newcommand{\fDSG}{\mathcal{DSG}}

\newcommand{\afPDS}{\widehat{\mathcal{PDS}}}

\newcommand{\afIPDS}{\widehat{\mathcal{IPDS}}}

\renewcommand{\Alphabet}{\Sigma}

\DeclareMathOperator*{\pdedge}{\rightarrowtail}

\newcommand{\biedge}[2]{#1 \mathrel{\rightarrowtail} #2}

\newcommand{\DSStates}{S}
\newcommand{\DSEdges}{E}
\newcommand{\DSFrames}{\StackAlpha}
\newcommand{\dsframe}{\stackchar}
\newcommand{\dsstate}{s}

\newcommand{\mkDSG}{\mathcal{F}}

\newcommand{\fnet}[1]{\lfloor #1 \rfloor}
\newcommand{\fstackify}[1]{\lceil #1 \rceil}

\newcommand{\ecg}{$\epsilon$-closure graph}

\DeclareMathOperator*{\RPDTrans}{{\longmapsto\!\!\!\!\!\!\!\longrightarrow}}

\DeclareMathOperator*{\areaches}{\rightarrowtriangle}

\newcommand{\aCollect}{{\hat{G}}}

\newcommand{\CallFrame} {\mathit{CallFrame}}
\newcommand{\aCallFrame}{\widehat{\mathit{CallFrame}}}

\newcommand{\HandlerFrame} {\mathit{HandlerFrame}}
\newcommand{\aHandlerFrame}{\widehat{\mathit{HandlerFrame}}}

\newcommand{\FieldAddr}{\mathit{FieldAddr}}
\newcommand{\aFieldAddr}{\sa{FieldAddr}}
\newcommand{\fa}{\mathit{fa}}
\newcommand{\afa}{\sa{fa}}

\newcommand{\RegAddr}{\mathit{RegAddr}}
\newcommand{\aRegAddr}{\sa{RegAddr}}
\newcommand{\ra}{\mathit{ra}}
\newcommand{\ara}{\sa{ra}}

\newcommand{\fp}{\s{fp}}
\newcommand{\objp}{\s{op}}
\newcommand{\aobjp}{\sa{op}}

\newcommand{\objv}{\mathit{ov}}
\newcommand{\aobjv}{\hat{\objv}}

\newcommand{\ObjectPointer}{\s{ObjectPointer}}
\newcommand{\aObjectPointer}{\widehat{\ObjectPointer}}

\newcommand{\FramePointer}{\s{FramePointer}}
\newcommand{\aFramePointer}{\widehat{\FramePointer}}

\newcommand{\FieldEval}{\mathcal{A}_{\mathcal{F}}}
\newcommand{\aFieldEval}{\hat\FieldEval}

\newcommand{\allocFP}{\s{allocFP}}
\newcommand{\allocOP}{\s{allocOP}}

\newcommand{\aallocFP}{\widehat{\allocFP}}
\newcommand{\aallocOP}{\widehat{\allocOP}}

\newcommand{\StmtOf}{{\mathcal{S}}}

\mathchardef\mhyphen="2D

\begin{document}
\title{Pushdown Exception-Flow Analysis of Object-Oriented Programs}
%
%
%

\author{Shuying Liang\inst{1} \and Matthew Might\inst{1} \and Thomas Gilray\inst{1} \and David Van Horn\inst{2} }


\institute{University of Utah, Salt Lake City, Utah, USA,\\
\email{\{liangsy, might, tgilray\}@cs.utah.edu}, 
\and
University of Northeastern University, \\
\email{dvanhorn@ccs.neu.edu}
}


\maketitle              

\begin{abstract}
  Statically reasoning in the presence of and about exceptions is challenging: 
exceptions worsen the well-known mutual recursion between data-flow and
control-flow analysis.
The recent development of pushdown control-flow analysis for the $\lambda$-calculus
hints at a way to improve analysis of exceptions: a pushdown stack can
precisely match catches to throws in the same way it matches returns to calls.
This work generalizes pushdown control-flow analysis 
to object-oriented programs and to exceptions.
Pushdown analysis of exceptions improves precision over the next best analysis, Bravenboer and Smaragdakis's Doop,
by orders of magnitude.
By then generalizing  abstract garbage collection to
object-oriented programs,
we reduce analysis time by half over pure pushdown analysis.
We evaluate our implementation for Dalvik bytecode on standard benchmarks as
well as several Android applications.

\end{abstract}

\section{Introduction}

Exceptions are not exceptional enough.
Thrown exceptions---or the possibility thereof---pervade the control-flow
structure of modern object-oriented programs.
A static analyzer grappling with
Java must concede that even
innocent-looking expressions like 
\begin{code}
 x / in.read() 
\end{code}
could throw four exceptions: 
{\tt ArithmeticException} (for divide by 0);
{\tt IOException} (for reading);
{\tt NullPointerException} (for dereferencing {\tt in});
and technically even
{\tt MethodNotFoundException} (if
the {\tt read} method
was removed after this file was compiled).

To make sense of a program, a static analyzer must exploit data-flow
information to rule out exceptions (such as {\tt NullPointerException} and {\tt
MethodNotFoundException} in the prior expression).
Yet, precise data-flow information \emph{requires} a precise analysis of
exceptions.
Co-analyzing data- and exception-flow is essential for precision.
Yet, even then, many exceptions (like {\tt IOException} or {\tt
ArithmeticException} in the prior expression) cannot be ruled out statically.
It is critical to precisely match catchers to throwers.

Exception-flow fundamentally follows the structure of the program stack at
run-time.
Because the stack can grow without bound, traditional analysis regimes like
$k$-CFA~\cite{mattmight:Shivers:1991:CFA} and its many variants implicitly or
explicitly finitize the stack during abstraction.
In effect, analyzers carve up dynamic return points and exception-handling
points among a finite number of abstract return contexts.
When two dynamic return points map to the same abstract context, the analyzer loses
the ability to distinguish them.
This confusion 
is a control-flow analog of the classic data-flow value merging problem.

To ground this discussion, 
consider the following Java fragment:
\begin{code}
try \{
   maybeThrow() ; // Call 1
\} catch (Exception e) \{
   System.err.println("Got an exception") ; // Handler 1
\}
maybeThrow() ; // Call 2\end{code}
Under a monovariant abstraction like 0CFA~\cite{mattmight:Shivers:1991:CFA}, where the distinction between different invocations of the same procedure are lost, it will seem as though exceptions
thrown from {\tt Call 2} can be caught by {\tt Handler 1}.

The fundamental problem with the analysis of exceptions is that the abstract
program stack is finite.
Our message is that pushdown analysis, which does not finitize the program
stack, is critical for precise analysis of exception-handling, yet it remains
computable.
Simply switching to pushdown analysis yields orders of magnitude improvements
in precision over \texttt{Doop}~\cite{mattmight:Bravenboer:2009:Exceptions},
the current state of the art exception-flow analysis.

Spotting an easy opportunity to improve running time, 
we reduce the state-space via abstract garbage collection~\cite{mattmight:Might:2006:GammaCFA}.
We then further improve running time and precision by
combining abstract garbage collection with live-range analysis.
In the end, we cut the time cost of pushdown exception-flow analysis by half.

Our implementation for Java (which targets the Dalvik virtual machine for Android)
 is publicly available:
\begin{center}
\verb+https://github.com/shuyingliang/pushdownoo+
\end{center}

\subsection{Contributions}

We make several contributions:

\begin{enumerate}

\item The first  application of the abstracting abstract
machines (AAM) methodology~\cite{mattmight:VanHorn:2010:Abstract} to create a
static analyzer for Java.

\item A pushdown flow analysis for precisely co-analyzing data-, control- and
exception-flow.

\item An empirical evaluation demonstrating two orders of magnitude of 
precision improvement over the current best 
analysis for exception-flow within reasonable analysis time.

\end{enumerate}

\section{The setting: An object-oriented bytecode} \label{sec:syntax}

In this section, we define an object-oriented bytecode
language 
closely modeled on the Dalvik virtual machine
to which Java applications for Android are compiled.
Subsequent sections develop our analysis for this language.

\subsection{Syntax}

The syntax of the bytecode language is given in Figure~\ref{fig:oo-syntax}.
Statements encode individual actions for the machine;
atomic expressions encode atomically computable values;
and complex expressions encode expressions with possible 
non-termination or side effects.
There are four kinds of names:
$\syn{Reg}$ for registers, $\syn{ClassName}$ for class names
$\syn{FieldName}$ for field names
and $\syn{MethodName}$ for method names.
There are two special register names:
$\syn{ret}$, which holds the return value of the last function called,
and $\syn{exn}$, which holds the most recently thrown exception.

The syntax is largely usual for an Java-like bytecode, but let us
explain the statements related to exceptions in
$\mathit{method}\mhyphen\mathit{def}$ in more detail:

\begin{itemize}
\item $(\syn{throws}~\mathit{class}\mhyphen\mathit{name} \dots)$
indicates that a  method makes a
\texttt{throws} declaration.
\item  $(\syn{push}\mhyphen\syn{handler} ~\mathit{class}\mhyphen\mathit{name}
~\mathit{label})$ pushes a handler frame on the stack.
The frame will catch exceptions of type $\mathit{class}$ 
and divert execution to $\mathit{label}$.

\item
$(\syn{pop}\mhyphen \syn{handler})$
pops the top-most handler frame off the stack.

\end{itemize}

\begin{figure}
 \begin{align*}
\mathit{program} &\produces 
\mathit{class} \mhyphen \mathit{def}\; \ldots
\\
\mathit{class} \mhyphen \mathit{def} &\produces (\mathit{attribute}\;\ldots~ \syn{class} ~\mathit{class}\mhyphen\mathit{name} ~\syn{extends}  ~\mathit{class}\mhyphen\mathit{name} 
\\
&\;\;\;\;\;\;\;\; (\mathit{field}\mhyphen{\mathit{def}}  \dots)\; (\mathit{method}\mhyphen\mathit{def} \dots))
\\
\mathit{field}\mhyphen\mathit{def} &\produces  (\syn{field}~\mathit{attribute}\;\ldots~\mathit{field}\mhyphen\mathit{name}~\mathit{type})
\\
\mathit{method}\mhyphen\mathit{def} \in \syn{MethodDef} &\produces  (\syn{method}~\mathit{attribute}\;\ldots~\mathit{method}\mhyphen\mathit{name}~(\mathit{type} \dots)~\mathit{type} 
\\
&\;\;\;\;\;\;\;\;
(\syn{throws}~\mathit{class}\mhyphen\mathit{name} \dots)
\;
(\syn{limit} ~n)
\; 
\stmt\; \ldots)
\\
\stmt \in \syn{Stmt}&\produces (\syn{label} ~\mathit{label}) \opor (\syn{nop}) \opor (\syn{line} ~\mathit{int}) \opor  (\syn{goto} ~\mathit{label}) 
\\
&\;\;\opor\;\;  (\syn{if}\; \aexpr ~(\syn{goto}~\mathit{label})) \opor (\syn{assign}~\mathit{\reg}\; [\aexpr \opor \cexpr]) \opor (\syn{return} ~\aexpr) 
\\
&\;\;\opor\;\; (\syn{field}\mhyphen\syn{put} ~\aexpr_o ~\mathit{field}\mhyphen\mathit{name} ~\aexpr_v) \opor (\syn{field}\mhyphen\syn{get}\; \$\reg ~\aexpr_o ~\mathit{field}\mhyphen\mathit{name})
\\
&\;\;\opor\;\; (\syn{push}\mhyphen\syn{handler} ~\mathit{class}\mhyphen\mathit{name} ~\mathit{label}) \opor (\syn{pop}\mhyphen \syn{handler})
\opor (\syn{throw}~\aexpr) 
\\
\aexpr \in \syn{AExp} &\produces \syn{this} \opor \syn{true} \opor \syn{false} \opor \syn{null} \opor \syn{void} \opor \mathit{\reg} \opor \mathit{int} 
\\
&\;\;\opor\;\;  (\mathit{atomic}\mhyphen\mathit{op} ~\aexpr \dots \aexpr)  \opor \syn{instance}\mhyphen\syn{of}(\aexpr, \mathit{class}\mhyphen\mathit{name}) 
\\
\cexpr &\produces (\syn{new}\; \mathit{class}\mhyphen\mathit{name}) \opor (\mathit{invoke}\mhyphen\mathit{kind} ~(\aexpr\dots\aexpr)\; (\mathit{type}_0\;\dots\;\mathit{type}_n))
\\
\mathit{invoke}\mhyphen\mathit{kind} &\produces \syn{invoke}\mhyphen\syn{static}  \opor \syn{invoke}\mhyphen\syn{direct} \opor \syn{invoke}\mhyphen\syn{virtual}  \opor \syn{invoke}\mhyphen\syn{interafce}  \opor \syn{invoke}\mhyphen\syn{super}
\\
\mathit{type} &\produces ~\mathit{class}\mhyphen\mathit{name} \opor \syn{int} \opor \syn{byte} \opor \syn{char} \opor \syn{boolean}
\\
\mathit{attribute} &\produces \syn{public} \opor \syn{private} \opor \syn{protected} \opor \syn{final} \opor \syn{abstract} 
\text.
\end{align*}
\caption{An object-oriented bytecode adapted from the Android specification~\cite{local:androidbytecode:url}.}


\label{fig:oo-syntax}
\end{figure}

With respect to a given program, we assume a syntactic metafunction
$\StmtOf : \syn{Label} \to \syn{Stmt}^*$, which maps a label to the
sequence of statements that start with that label.

\subsection{Concrete semantics} \label{sec:conc-sem}

Interpretation of bytecode programs is defined in terms of a
CESK-style machine model.  States of this machine consist of a series
of statements, a frame pointer, a heap, and a stack.  The evaluation
of a program is defined as the set of machine configurations reachable
by machine transitions from the initial program.  Formally, the
evaluation function,
$ \Eval : \syn{Stmt^*} \to \Pow{\s{Conf}}$,
is defined as:
  \begin{align*}
 \Eval(\vec{s}) = \setbuild{ \conf }{ \Inject(\vec{s}) \To^* \conf }\text.
\end{align*}
This function injects, using $\Inject : \syn{Stmt^*} \to \s{Conf}$, an
initial program sequence into an initial machine configuration.  From
this initial configuration, evaluation is defined by the set of
configurations reached by the reflexive, transitive closure of the
machine transition relation, $(\To) \subseteq \s{Conf} \times
\s{Conf}$.  The next section describes the details of machine
configurations; the subsequent section defines the machine
transition relation, $(\To)$.

\subsection{Concrete configuration-space} \label{subsec:conc-ss}

Figure \ref{fig:conc-conf-space} presents the machine's concrete
configuration-space.  The machine has an explicit stack, which under
structural abstraction will become the stack component of a pushdown
system.  The stack contains not only call frames, but also mini-frames
for exception handlers.
The $\FramePointer$ is the environmental
component of the machine: by pairing the frame pointer with a
register name, it forms the address of its value in the store.

 \begin{figure}
\begin{align*}
 \conf \in \s{Conf} &= \syn{Stmt^*} \times \s{FramePointer} \times \s{Store} \times \s{Kont} && \text{[configurations]}
\\
 \store \in \s{Store} &= \s{Addr} \to \s{Val} && \text{[stores]}
 \\
\addr \in  \s{Addr} &=  \RegAddr   +  \FieldAddr&& \text{[addresses]}
 \\
  \ra \in \RegAddr &= \s{FramePointer}  \times \syn{Reg}  
  \\
 \fa \in \FieldAddr &= \s{ObjectPointer} \times {\syn{FieldName} }
\\
\cont \in \s{Kont} &= \s{Frame}^* && \text{[continuations]}
\\
\phrame \in \s{Frame} &= \CallFrame + \HandlerFrame 
\\
\callframe \in \CallFrame &\produces \mathbf{fun}(\fp,\vec{s})
\\
\exnframe \in \HandlerFrame &\produces \mathbf{handle}(\mathit{class\text{-}name},
\mathit{label})
\\
\den \in \s{Val} &= \s{ObjectValue} + \mathit{String} + \mathcal{Z} && \text{[values]}
\\
\objv \in \s{ObjectValue} &= \mathit{ObjectPointer} \times \syn{ClassName}
\\
fp \in \s{FramePointer} &\text{ is an infinite set of frame pointers} && \text{[frame pointers]}
\\ 
\objp \in \s{ObjectPointer}&\text{ is an infinite set of object pointers} && \text{[object pointers]}
\text.
 \end{align*}
\caption{The concrete configuration-space.}
\label{fig:conc-conf-space}
 \end{figure}

The initial configuration consists of the program, the initial frame
pointer, an empty heap, and an empty stack:
\begin{equation*}
\conf_0 = \Inject(\vec{s}) = (\vec{s}, \fp_{0}, [], \vect{})\text.
\end{equation*}

\subsection{Concrete transition relation} \label{subsec:conc-tr}

In this section, we describe the essential cases of the $(\To)$
relation, which deal with objects and exceptions.  The remaining cases
are in Appendix~\ref{appendix:simple-conc-rules}.

The machine relies on helper functions for evaluating atomic
expressions, looking up field values, and allocating memory:
 \begin{itemize}

  \item $\ArgEval : \syn{AExp} \times \s{FramePointer} \times \s{Store} \parto
  \s{Val}$ evaluates atomic expressions:
  \begin{align*}
    \ArgEval(\mathit{\reg},\fp,\store) &= \store(\fp,\reg) && \text{[variable look-up]}
    \text.
    \end{align*}

\item
 $\FieldEval : \syn{AExp} \times \s{FramePointer} \times \s{Store} \times \syn{FieldName} \parto \s{Val}$ looks up fields:
  \begin{align*}
    \FieldEval(\aexpr,\fp,\store, \mathit{field\mhyphen name}) &=  \store(\op,\mathit{field\mhyphen name}) && \text{[field look-up]}
    \\ 
    \text{where}~(\op, class\mhyphen name) &=  \ArgEval(\aexpr, \fp, \store )
    \text.
    \end{align*}

\end{itemize}

\paragraph{Allocation}
$\FramePointer$ and $\ObjectPointer$ 
determine 
  addresses for $\RegAddr$ and $\FieldAddr$
  respectively.
We need to specify how to allocate these pointers:
\begin{itemize}

\item
  $\allocFP : \s{Conf} \to \FramePointer$ 
  chooses a fresh frame pointer for newly invoked method. 

\item
$\allocOP : \s{Conf} \to \ObjectPointer$,
 allocates a  fresh object pointer in the instantiation site.

\end{itemize}
For the sake of defining a concrete semantics,
these could allocate increasingly larger natural numbers.
  Under abstraction, these parameters provide the knob to tune the
  polyvariance, context-sensitivity and object-sensitivity 
  of the resulting analysis.

\subsubsection{New object creation}

Creating an object allocates a new object pointer,
creates a fresh address for the register 
and initializes the fields:
\begin{align*}
    \overbrace{
      (\sembr{(\syn{assign}\;\reg\; (\syn{new}\; \mathit{class}\mhyphen\mathit{name})) :  \vec{s}}, \fp, \store,  \cont)
     }^{\conf}
    &\To 
       (\vec{s}, \fp, \store'',  \cont)
       \text{, where }
     \\
    \mathit{op} &= \allocOP(\conf)
   \\
   \store' &= \store[(\fp, \mathit{\reg}) \mapsto  (\mathit{op}, \mathit{class}\mhyphen\mathit{name})]
   \\
  \store'' &= \mathit{initObject}(\store', \mathit{class}\mhyphen\mathit{name})
  \text{.}
    \end{align*}
The  helper function,     $\mathit{initObject}: \Store \times \syn{ClassName} \rightharpoonup \Store$,  
  initializes the field addresses in the provided store.

\subsubsection{Instance field reference/update}
Referencing a field gets the object pointer
and then grabs the field value as an offset:
\begin{align*}
  (\sembr{(\syn{field}\mhyphen\syn{get}\; \reg ~\aexpr_o ~\mathit{field}\mhyphen\mathit{name}) : \vec{s}}, \fp, \store,\cont)
  &\To
    (\vec{s},\fp,\store', \cont)
 \text{, where}
 \\
 \store' = \store[ (\fp, \reg)&\mapsto \FieldEval(\ae_o, \fp, \store, \mathit{field}\mhyphen\mathit{name} ) ] \text.
  \end{align*}
Updating a field grabs the object, extracts the object pointer
and updates the associated field in the store:
\begin{align*}
    (\sembr{(\syn{field}\mhyphen\syn{put}~\ae_o~\mathit{field}\mhyphen{name} ~\ae_v) : \vec{s}}, \fp, \store, \cont)
  &\To
    (\vec{s},\fp,\store', \cont)
 \text{, where}
 \\
 \store' = \store[(\op, \mathit{field}\mhyphen{name}) &\mapsto \ArgEval(\ae_v, \fp, \store)] 
 \\
 (\op, \mathit{class}\mhyphen{name}) &= \ArgEval(\ae_o, \fp, \store)
 \text.
  \end{align*}

\subsubsection{Method invocation}

Method invocation involves all four components of the
machine.
Since the language supports inheritance, method resolution requires a traversal
of the class hierarchy.  This traversal
is not of interest, so we focus on the
helper function that performs method application:
$\mathit{applyMethod}$. %
The function $\mathit{applyMethod}$ takes a method definition,
arguments,
a frame pointer,
a store and a continuation and produces the next configuration:
\begin{align*}
\mathit{applyMethod} : \syn{MethodDef} \times \syn{AExp}^* \times \FramePointer \times \Store \times \s{Kont} \rightharpoonup
\s{Conf}\text.
  \end{align*}
It looks up the values of the arguments, binds them to the formal parameters of the method, creates a new frame pointer and  a new continuation:
\begin{align*}
 \mathit{applyMethod}(m, \vec{\ae}, \fp, \store, \cont) &= (\vec{\stmt}, \fp', \store', (\fp, \vec{s}) : \cont), \text{where} ~\fp' ~\text{is fresh},
\\
\store' &= \store[(\fp', \reg_i) \mapsto \ArgEval(\ae_i, \fp, \store)]\text.
\end{align*}
Finally, the transition looks up the method
$m$ and then passes it to $\mathit{applyMethod}$:
\begin{align*}
    (\sembr{ (\mathit{invoke}\mhyphen\mathit{kind} ~(\aexpr_0\dots\aexpr_n) (\mathit{type}_0\dots\mathit{type}_n))} :  \vec{s}, \fp, \store, \cont)
  &\To 
     \mathit{applyMethod}(m, \vec{\aexpr}, \fp,\store,\cont)
 \text.
  \end{align*}

\subsubsection{Procedure return}
Returning a value restores the caller's context and puts the return value in
the dedicated return register,  $\syn{ret}$.
\begin{align*}
      (\sembr{(\syn{return}~\aexpr) :  \vec{s}}, \fp, \store, \mathbf{fun}(\fp',\vec{s}') : \cont)
    &\To 
       (\vec{s}', \fp', \store',  \cont)
   \text{, where}
   \\
   \store' & = \store[(\fp', \syn{ret})\mapsto \ArgEval(\aexpr, \fp, \store)]\text.
    \end{align*}
If a $\HandlerFrame$ is on top of the stack, 
the transition will pop it
without changing any other part of the state:
 \begin{align*}
      (\sembr{(\syn{return} ~\aexpr)} :  \vec{s}, \fp, \store,
\mathbf{handle}(\mathit{class}\mhyphen\mathit{name}~\mathit{label}) : \cont)
  &\To 
       (\sembr{(\syn{return} ~\aexpr)}:  \vec{s}, \fp,\store
       , \cont) 
 \text{.}
  \end{align*}

\subsubsection{Pushing and popping exception handlers} 
Pushing and popping exception handlers is straightforward:
\begin{align*}
    (\sembr{(\syn{push}\mhyphen\syn{handler}~\mathit{class}\mhyphen\mathit{name}~\mathit{label})} :  \vec{s}, \fp, \store, \cont)
  &\To 
     (\vec{s},\fp,\store,\mathbf{handle}(\mathit{class}\mhyphen\mathit{name}~\mathit{label}) : \cont)
 \text,
  \end{align*}
 \begin{align*}
    (\sembr{(\syn{pop}\mhyphen\syn{handler})} :  \vec{s}, \fp, \store, \mathbf{handle}(\mathit{class}\mhyphen\mathit{name}~\mathit{label}) : \cont)
  &\To 
     (\vec{s},\fp,\store, \cont)
 \text.
  \end{align*}

\subsubsection{Throwing and catching exceptions}
The {\tt{throw}} statement peels away layers of the stack until it finds a matching exception handler:
 \begin{align*}
    (
      \sembr{(\syn{throw}~\aexpr)} :  \vec{s}, \fp, \store
      ,\cont)
  &\To
  \mathit{handle}(\ae, \vec{s}, \fp, \store, \cont)
 \text,
  \end{align*}
where the function $\mathit{handle}
: \syn{AExp} \times \syn{Stmt^*} \times \mathit{FramePointer} \times \s{Store} \times \s{Kont}$
$\rightharpoonup \mathit{Conf}$.
does the peeling.
If  a matching handler is found, that is, $\mathit{class}\mhyphen\mathit{name}$ is a subclass of  $\mathit{class}\mhyphen\mathit{name'}$,
where $( \objp, \mathit{class}\mhyphen\mathit{name}) = \ArgEval(\ae, \fp, \store)$ 
and $\mathit{class}\mhyphen\mathit{name}'$ is from the top $\HandlerFrame$,
the execution flow jumps to code block of the handler:
\begin{gather*}
 \mathit{handle}(\ae, \vec{s}, \fp, \store,  \mathbf{handle}(\mathit{class}\mhyphen\mathit{name'}~\mathit{label}) : \cont')
 = \\
   (\StmtOf(\mathit{label}), \fp, \store[(\fp, \syn{exn}) \mapsto( \objp, \mathit{class}\mhyphen\mathit{name})], \cont') 
 \text.
\end{gather*}
The last thrown exception object value will be put in 
 the dedicated exception register $\syn{exn}$.
 
If the exception type does not match or it's a call frame, 
then $\mathit{handle}$ transits to a configuration 
with the control state unchanged but with the
 top frame popped:
  \begin{align*}
  \mathit{handle}(\ae, \vec{s}, \fp, \store,  \mathbf{handle}(\mathit{class}\mhyphen\mathit{name'}~\mathit{label}) : \cont')
  &=
      (\sembr{(\syn{throw}~\aexpr)} :  \vec{s}, \fp, \store
      , \cont')
\\
  \mathit{handle}(\ae, \vec{s}, \fp, \store,  \mathbf{fun}(\fp',\vec{s}') : \cont')
  &=
      (\sembr{(\syn{throw}~\aexpr)} :  \vec{s}, \fp, \store
      , \cont')
  \text.
    \end{align*}
The abstraction of these ``multi-pop'' transition relations
will require modification of the 
algorithm used for control-state reachability
(Section~\ref{subsec: dcg-exn}).

\section{Pushdown abstract semantics}
\label{sec:pdexflowoo}

With the concrete semantics in place,
it is time to abstract them into an analysis.
To achieve a pushdown analysis, we abstract less than we normally would. 
Specifically, we
conduct a structural abstraction of the concrete state-space
and leave the stack height unbounded
rather that thread frames through the heap.

\subsection{Abstract semantics}

Abstract interpretation is defined in terms of a structural
abstraction of the machine model of Section~\ref{sec:syntax}.  The
evaluation of a program is defined as the set of \emph{abstract}
machine configurations reachable by an abstraction of the machine
transitions relation. Largely, abstract evaluation, $\aEval :
\syn{Stmt^*} \to \PowSm{\sa{Conf}}$, mimics its concrete counterpart:
\begin{equation*}
  \aEval(\vec{s}) = \setbuild{ \aconf }{ \aInject{(\vec{s}) \aTo^* \aconf }} 
  \text.
\end{equation*}
Abstract evaluation is defined by the set of configurations reached by
the reflexive, transitive closure of the $(\aTo)$ relation, which
abstracts the $(\To)$ relation.

\subsection{Abstract configuration-space}

Figure~\ref{fig:abs-conf-space} details the abstract configuration-space. 
We assume the natural element-wise, point-wise and member-wise
lifting of a partial order across this state-space.

 \begin{figure}
\begin{align*}
 \aconf \in \sa{Conf} &= \syn{Stmt^*} \times \sa{FramePointer} \times \sa{Store} \times \sa{Kont} && \text{[configurations]}
\\
 \astore \in \sa{Store} &= \sa{Addr}  \parto   \sa{Val} && \text{[stores]}
 \\
\aaddr \in  \sa{Addr} &=  \aRegAddr  + \aFieldAddr&& \text{[addresses]}
 \\
  \ara \in \aRegAddr &= \sa{FramePointer}  \times \syn{Reg}  
  \\
 \afa \in \aFieldAddr &= \sa{ObjectPointer} \times {\syn{FieldName} }
\\
\acont \in \sa{Kont} &= \sa{Frame}^* && \text{[continuations]}
\\
\aphrame \in \sa{Frame} &= \aCallFrame + \aHandlerFrame    &&   \text{[stack frames]}
\\
\acallframe \in \aCallFrame &\produces \mathbf{fun}(\afp,\vec{s})
\\
\aexnframe \in \aHandlerFrame &\produces \mathbf{handle}(\mathit{class\text{-}name},\mathit{label})
\\
\aden \in \sa{Val} &= \Pow{ \sa{ObjectValue} +  \sa{String} +  \sa{\mathcal{Z}}} && \text{[abstract values]}  
 \\
 \aobjv \in \sa{ObjectValue} &= \aObjectPointer \times \syn{ClassName}
\\
\afp \in \aFramePointer &\text{ is a finite set of frame pointers} && \text{[frame pointers]}
\\ 
\aobjp \in \aObjectPointer&\text{ is a finite set of object pointers} && \text{[object pointers]}
\text.
 \end{align*}
\caption{The abstract configuration-space.}
\label{fig:abs-conf-space}
 \end{figure}

To synthesize the abstract state-space,
we force frame pointers and object pointers (and thus addresses) to be a finite set, 
but crucially, we leave the stack untouched. 
When we compact the set of addresses into a finite set, 
the machine may run out of addresses to allocate, and when it does, 
the pigeon-hole principle will force multiple abstract values to reside at the same address. 
As a result, we have no choice but to force the range of the $\aStore$ to become a power set 
in the abstract configuration-space.

\subsection{Abstract transition relation} \label{sec:abs-transition-rules}

The abstract transition relation has components analogous to those from the concrete semantics:

\begin{itemize}

\item
$\Inject : \syn{Stmt^*} \to \sa{Conf}$ injects an sequence of instructions into
                                              a configuration:
 \begin{equation*}
\aconf_0 = \aInject(\vec{s}) = (\vec{s}, \afp_0, [], \vect{})
  \text.
  \end{equation*}

\item
$\aArgEval : \syn{AExp} \times \aFramePointer \times \aStore \rightharpoonup
  \sa{Val}$ evaluates atomic expressions:
  \begin{align*}
    \aArgEval(\reg,\afp,\astore) &= \store(\afp,\reg) && \text{[variable look-up]}
    \text.
    \end{align*}

\item
$ \aFieldEval : \syn{AExp} \times  \aFramePointer \times \aStore \times \syn{FieldName} \rightharpoonup \sa{Val}$
looks up fields:
  \begin{align*}
    \aFieldEval(\aexpr,\afp,\astore, \mathit{field}\mhyphen \mathit{name}) &=  \bigjoin \astore(\aobjp,\mathit{field}\mhyphen \mathit{name}) && \text{[field look-up]}
    \\ 
    ~\text{where} ~(\aobjp, \mathit{class}\mhyphen \mathit{name}) &\in  \aArgEval(\aexpr, \afp, \astore )
    \text.
    \end{align*}

\end{itemize}
      Because there are an infinite number of abstract configurations, a
      na\"ive implementation of the $\aEval$ function may not terminate.

Appendix~\ref{sec:polyvariance} discusses abstractions of $\allocFP$ and
$\allocOP$ that allow the selection of different analyses such as $k$-CFA or
polymorphic splitting.

 The rules for the abstract transition relation $(\aTo) \subseteq \sa{Conf} \times
    \sa{Conf}$ 
    largely mimic the structure of the concrete relation $(\To)$.
The biggest difference is that the structural abstraction 
forces the abstract transition to become nondeterministic.
 We detail these rules below and illustrate the 
 differences from its concrete counterpart. 
 Again, we only cover rules involving objects and exceptions. 
 Appendix~\ref{appendix-simple-abs-trs} contains the remaining rules.

\subsubsection{\textbf{New object creation}} 
Creating an object allocates a potentially non-fresh object pointer
and joins the newly initialized object into that store location:
\begin{align*}
    \overbrace{(\sembr{(\syn{assign}\; \reg\; (\syn{new}\;\mathit{class}\mhyphen\mathit{name})) :  \vec{s}}, \afp, \astore,  \acont)}^{\aconf}
    &\To 
    (\vec{s}, \afp, \astore'',  \acont), 
   \\
   \aobjp' = \aallocOP(\aconf)
   \\
   \astore' = \astore \join [(\afp, \mathit{\reg}) &\mapsto  ( \aobjp', \mathit{class}\mhyphen\mathit{name})]
   \\
  \astore'' = \widehat{\mathit{initObject}}(\astore', &\mathit{class}\mhyphen\mathit{name})
  \text{,}
    \end{align*}
where the  helper 
$\widehat{\mathit{initObject}}: \aStore \times \syn{ClassName} \rightharpoonup \aStore$ initializes fields. 
  
  \subsubsection{\textbf{Instance field reference/update}}
Referencing a field uses $\aFieldEval$ to evaluate the field values and join the store for destination register:
 \begin{align*}
     (\sembr{(\syn{field}\mhyphen\syn{get}\; \reg ~\aexpr_o ~\mathit{field}\mhyphen\mathit{name}) : \vec{s}}, \afp, \astore,\acont)
   &\aTo
     (\vec{s},\afp,\astore', \acont)
  \text{, where}
  \\
  \astore' = \astore \join [ (\afp, \reg)&\mapsto \aFieldEval(\aexpr_o, \afp, \astore, \mathit{field}\mhyphen\mathit{name} ) ] 
  \text.
   \end{align*}
   Updating a field first finds the abstract object values from the
   store, extracts its object pointer from each of all the
   possible values, then pairs this object pointer with the field name
   to get the field addresses, and finally \textit{joins} the extensions
   to the store:
 \begin{align*}
     (\sembr{(\syn{field}\mhyphen\syn{put}~\aexpr_o~\mathit{field}\mhyphen{name} ~\aexpr_v) : \vec{s}}, \afp, \astore, \acont)
   &\aTo
     (\vec{s},\afp,\astore', \acont)
  \text{, where}
  \\
  \astore' = \astore \join [(\aobjp, \mathit{field}\mhyphen{name}) &\mapsto \aArgEval(\aexpr_v, \afp, \astore)] 
  \\
  (\aobjp, \mathit{class}\mhyphen{name}) &\in \aArgEval(\aexpr_o, \afp, \astore)\text.
   \end{align*}

\subsubsection{\textbf{Method invocation}}
Like the concrete semantics, method invocation also involves all four components of the machine.
The main difference is that, for non-static methods invocation, there can be a \emph{set} of possible objects that are invoked, 
rather than only one as in its concrete counterpart. 
 This also means that there could be multiple method definitions resolved for each object.
For each such method $m$: 
 \begin{align*}
  \overbrace{
    (\sembr{ (\mathit{invoke}\mhyphen\mathit{kind} ~(\aexpr_0\dots\aexpr_n)\; (\mathit{type}_0\dots\mathit{type}_n))} :  \vec{s}, \afp, \astore, \acont)
    }^{\aconf}
  &\aTo 
     \widehat{\mathit{applyMethod}}(m, \vec{\aexpr},
     \afp,\astore,\acont)
\text,
  \end{align*}
where,
\begin{align*}
 \widehat{\mathit{applyMethod}}(m, \vec{\aexpr}, \afp, \astore, \acont) &= (\vec{\stmt}, \afp', \astore', (\afp, \vec{s}) : \acont), \text{where}
\\
\afp' &=  \aallocFP(\aconf)
\\
 \astore' &= \astore \join [(\afp', \reg_i) \mapsto \aArgEval(\aexpr_i, \afp, \astore)]\text.
\end{align*}

\subsubsection{\textbf{Procedure return}}
Procedure return pops off the top-most $\mathbf{fun}$ frame:
\begin{align*}
      (\sembr{(\syn{return}~\aexpr) :  \vec{s}}, \afp, \astore,  \mathbf{fun}(\afp',\vec{s}') : \acont)
    &\aTo 
       (\vec{s}', \afp', \astore',  \acont)
   \text{, where}
   \\
   \astore' &= \astore\join [(\afp', \syn{ret})\mapsto \aArgEval(\aexpr, \afp, \astore)]
   \text.
    \end{align*}
If the top frame is a $\mathbf{handle}$ frame, 
the abstract interpreter pops until the top-most frame is 
a $\mathbf{fun}$ frame:
 \begin{align*}
      (\sembr{(\syn{return} ~\aexpr)} :  \vec{s}, \afp, \astore
      ,
      \mathbf{handle}(\mathit{class}\mhyphen\mathit{name}~\mathit{label}) : \acont)
  &\aTo 
     (\sembr{(\syn{return} ~\aexpr)}:  \vec{s}, \afp,\astore
     , \acont)
 \text{.}
  \end{align*}

  \subsubsection{\textbf{Pushing and popping handlers}}
  Handlers push and pop as expected:
   \begin{align*}
      (\sembr{(\syn{push}\mhyphen\syn{handler} ~\mathit{class}\mhyphen\mathit{name}~\mathit{label})} :  \vec{s}, \afp, \astore, \acont)
    &\aTo 
       (\vec{s},\afp,\astore,\mathbf{handle}(\mathit{class}\mhyphen\mathit{name}~\mathit{label}) : \acont)
    \end{align*}
   \begin{align*}
      (\sembr{(\syn{pop}\mhyphen\syn{handler})} :  \vec{s}, \afp, \astore, \mathbf{handle}(\mathit{class}\mhyphen\mathit{name}~\mathit{label}) : \acont)
    &\aTo 
       (\vec{s},\afp,\astore, \acont)
   \text.
    \end{align*}
  \subsubsection{\textbf{Throwing and catching exceptions}}
  The throw statement peels away layers of the stack until it finds a matching exception handler:
   \begin{align*}
      (\sembr{(\syn{throw}~\aexpr)} :  \vec{s}, \afp, \astore
      ,\acont)
    &\aTo
    \widehat{\mathit{handle}}(\aexpr, \vec{s}, \afp, \astore, \acont)
   \text,
    \end{align*}
  where the function 
  $\widehat{\mathit{handle}} : \syn{AExp} \times \syn{Stmt^*} \times  \aFramePointer  \times \aStore \times \sa{Kont}$
  $\rightharpoonup \sa{Conf}$
  behaves like its concrete counterpart when the top-most frame is a
  compatible handler:
\begin{align*}
   &\widehat{\mathit{handle}}(\aexpr, \vec{s}, \afp, \astore,  \mathbf{handle}(\mathit{class}\mhyphen\mathit{name'}~\mathit{label}) : \acont')
   \\
   &\;\;\;\;= 
     (\StmtOf(\mathit{label}), \afp, \astore \join [(\afp, \syn{exn}) \mapsto( \aobjp, \mathit{class}\mhyphen\mathit{name})], \acont') 
   \text.
\end{align*}
Otherwise, it pops a frame:
\begin{align*}
    \widehat{\mathit{handle}}(\aexpr, \vec{s}, \afp, \astore,  \mathbf{handle}(\_,\_) : \acont')
    &=
     (\sembr{(\syn{throw}~\aexpr)} :  \vec{s}, \afp, \astore
   , \acont') 
\\
    \widehat{\mathit{handle}}(\aexpr, \vec{s}, \afp, \astore,  \mathbf{fun}(\_,\_) : \acont')
    &=
     (\sembr{(\syn{throw}~\aexpr)} :  \vec{s}, \afp, \astore
   , \acont') 
    \text.
\end{align*}

\section{The shift: From abstract CESK to pushdown systems} 

  In the previous section, we constructed an infinite-state abstract
    interpretation of the CESK-like machine to analyze exception flows 
    for object-oriented languages.
    The infinite-state nature of the abstraction makes it difficult 
    to answer static analysis questions:
    How do you compute the reachable states if there are an infinite number of them?
    Fortunately,   a shift in perspective reveals that the machine 
is in fact a pushdown system for which control-state reachability is decidable.

     If we take $\Stmt^* \times \sa{FramePointer} \times \sa{Store}$
     as the finite set of control states and
     $\sa{Kont}$ is the set of stacks,
     then it is immediately apparent that 
     the abstract semantics that we have created is a pushdown
     system.
     This is the object-oriented analog of Earl \etal's observation
     for the $\lambda$-calculus~\cite{Earl:2012:IPDCFA}.
    This shift permits the use of a control-state reachability
       algorithm in place of exhaustive search of the
       configuration-space.
       [Appendix Figure~\ref{fig:acesk-to-pds} defines the program-to-RPDS conversion function
                 $\afPDS : \syn{Stmt^*} \to \mathbb{RPDS}$ in detail.]

\subsection{Mini-evaluation}

In Table~\ref{tbl:result}, when we compare the resulting analysis to 
Bravenboer and Smaragdakis's 
finite-state analysis of exceptions~\cite{mattmight:Bravenboer:2009:Exceptions},
we find a solid improvement in precision, but a substantial slowdown in time.
This is not surprising: 
computing the reachable states in a pushdown system 
is cubic in the number of states.
In the next section, we improve the running time
by porting another powerful technique from abstract interpretation of the $\lambda$-calculus:
abstract garbage collection.

\section{Abstract garbage collection for objects} \label{sec:agc-oo1}

Abstract garbage collection is known to yield order-of-magnitude improvements
in precision, even as it drops run-times by cutting away false positives.
Adapting abstract garbage collection seemed like the right tool to fix 
the performance problem of the previous section.
We directly benefit from that line of work on the $\lambda$-calculus, which 
developed a class of
\emph{introspective} pushdown machines as a means of combining
pushdown analysis with abstract garbage collection~\cite{Earl:2012:IPDCFA}.
Introspective pushdown systems are pushdown systems that have read access
to the \emph{entire} stack during a transition.
Since the root set for garbage collection depends on the entire stack,
we need an introspective pushdown systems to use abstract garbage collection.
[Appendix~\ref{subsubsec: gc-ipds} formalizes
the injection into an introspective pushdown system.]

It's natural to think that the combined technique will benefit 
exception-flow analysis for object-oriented languages.
    However, as we shall demonstrate, 
    we must conduct a careful and subtle
    redesign of the abstract garbage collection machinery 
    for object-oriented languages
     to gain the promised analysis precision and performance.

 In the following, we present  how to adapt abstract garbage collection to work under 
abstract semantics defined in Section~\ref{sec:pdexflowoo}.
Abstract garbage collection discards unreachable elements from the
store.
It modifies the transition relation
to conduct a ``stop-and-copy'' garbage collection before each
transition.
To do so, we define a garbage collection function 
$\aCollect : \sa{Conf} \to \sa{Conf}$
on
configurations:
\begin{align*}
\aCollect(\overbrace{\vec{s},\afp,\astore,\acont}^{\aconf})
&= (\vec{s},\afp,\astore|\mathit{Reachable}(\aconf),\acont)
  \text,
  \end{align*}
  where the pipe operation $f|S$ yields the function $f$, but with
  inputs not in the set $S$ mapped to bottom---the empty set.
  The reachability function $\mathit{Reachable} : \sa{Conf} \to \PowSm{\sa{Addr}}$
  first computes the root set, and then the transitive closure of an
  address-to-address adjacency relation: 
  \begin{align*}
  \mathit{Reachable}(\overbrace{\vec{s},\afp,\astore,\acont}^\aconf) &=
\setbuild{ \aaddr }{ \aaddr_0 \in \mathit{Root}(\aconf)
  \text{ and }
  \aaddr_0  
    \mathrel{\areaches_\astore^*}
  \aaddr
}
\text,
  \end{align*}
  where the function $\mathit{Root} : \sa{Conf} \to 
  \PowSm{\sa{Addr}}$ 
  finds the root addresses:
  \begin{align*} 
  \mathit{Root}(\vec{s}, \afp,\astore,\acont) &=
 \{(\afp, r): (\afp,r) \in \mathit{dom}(\astore) \} \union
\mathit{StackRoot}(\acont)
  \text,
  \end{align*}

  The $\mathit{StackRoot} : \sa{Kont} \to \PowSm{\sa{Addr}}$ function
  finds roots down the stack. 
However,  only $\aCallFrame$ has the component to construct addresses, so we define a 
helper function $\hat{\mathcal{F}} : \sa{Kont} \to \aCallFrame^*$ 
to extract only $\aCallFrame$ out from the stack and skip over all the handle frames.
Now $\mathit{StackRoot} $ is defined as
 \begin{align*} 
  \mathit{StackRoot}(\acont)
  &= 
\{(\afp_i, r) : (\afp_i, r) \in \mathit{dom}(\astore) ~\text{and}~ \afp_i \in \hat{\mathcal{F}}(\acont)\}
  \text,
  \end{align*}
  and the relation:
  \[
  (\areaches) \subseteq \sa{Addr} \times \sa{Store} \times \sa{Addr}
  \]
  connects adjacent addresses: 
 \[
 \aaddr 
  \mathrel{\areaches_\astore} 
  \aaddr'
  \text{ iff there exists }
(\aobjp, \mathit{class}\mhyphen\mathit{name}) \in \astore(\aaddr)
\]
 \text{ such that }
  $\aaddr' \in \{(\aobjp,  \mathit{field}\mhyphen\mathit{name}): (\aobjp,  \mathit{field}\mhyphen\mathit{name}) \in \mathit{dom}(\astore)\}$.

  \subsubsection{Example runs with abstract garbage collection}

Table~\ref{tbl:to-lead-lra} presents the example results of running pushdown analysis with and without abstract garbage collection,
as described.
It shows that abstract garbage collection further improves the precision, but the effect  is not as large as we had predicted,
especially with respect of analysis time, where on functional programs, abstract garbage collection can bring order-of-magnitude 
reductions in both  imprecision and time.
The next section teases out the problem and develops a solution: combining abstract garbage collection 
with live range analysis.

\begin{table}
    \begin{tabular}{ c | c | c | c | c | c | c }
    \textbf{Benchmark} & \textbf{Opts} & \textbf{Nodes} & \textbf{Edges} &  \textbf{VarPointsTo} & \textbf{E-C links} & \textbf{Time(sec)} \\ \hline 
          \multirow{2}{*}{lusearch} & pdcfa & 91574  & 105154  &  (1423, 3) &  76  &  5520 
          \\ & +gc  & 26365  & 30426  &  (1086, 2) &  63 & 4800 
         \\
     \hline
    \end{tabular}    
     \caption{Example analysis result by (introspective) pushdown system.
     {\tt{VarPointsTo}} measures how many objects can a variable possibly points to; 
     it is   presented as a tuple $(a,b)$, where $a$ is the total entries, $b$ is the average 
         objects being invoked on;
     {\tt{E-C links}} is the number of pairs of  an instruction that can throw exceptions and a handler that can possibly handle the exception.
     These metrics are used by Fu, \etal~\cite{Fu:2005:rubust-java-server-apps}, and Bravenboer and
     Smaragdakis~\cite{Bravenboer:2009:Exceptions}.
     }
     \label{tbl:to-lead-lra}
\end{table}

\subsection{Live register analysis (LRA) for AGC} \label{subsec:lra}

Even though pushdown analysis with/without garbage collection
promises to increase analysis precision, 
the analysis time is not satisfying, 
as shown in Table~\ref{tbl:to-lead-lra}.
The benchmark $\syn{lusearch}$ with abstract garbage collection still takes more than an hour.
By manual inspection on some other benchmarks we have run, 
we find that 
in the register-based byte code, 
there are cases that the same register is reassigned multiple 
    times at different sites within a method. 
    Therefore, abstract object values are unnecessarily  ``merged'' together.
    The result is that unnecessary state space is explored and analysis time is prolonged.

    The direct adaptation of AGC to an object-oriented setting in Section~\ref{sec:agc-oo1}
    cannot collect these registers between uses.
    For object-oriented programs, we want to collect registers that are reachable, but not without
    an intervening assignment.

    As it turns out, the fix for this problem 
    is a classic data-flow analysis:
    live-register analysis (LRA).
        LRA can compute the set of registers that are \textit{alive} at each statement within a method.
The garbage collector can then more precisely collect each frame.

  Since LRA is well-defined in the literature~\cite{local:new-dragon}, we skip the formalization here, 
  but   
   the $\mathit{Root}$  is now modified to collect
      only \textit{living}  registers of the current statement $\Lives\{s_0\}$:
      \begin{align*} 
      \mathit{Root}(\vec{s}, \afp,\astore,\acont) &=
     \{(\afp, r'): (\afp,r') \in \mathit{dom}(\astore) ~\text{and}~ r' \in \Lives\{s_0\}  \} \union
    \mathit{StackRoot}(\acont)
      \text.
      \end{align*}
 
Section~\ref{sec:evaluation} presents the complete results 
running on the suite of the benchmarks based on 
the joint analysis (denoted as {\tt{+gc+lra}} in Table~\ref{tbl:result}).

\section{Extending pushdown reachability to exceptions} \label{sec:impl}
With the formalism in previous sections, 
it is not hard to translate the abstract semantics into working code.
We use the Dyck State Graph synthesis algorithm---a purely functional 
version of the Summarization algorithm---for 
computing reachable pushdown control states~\cite{Earl:2012:IPDCFA}.

\subsection{Synthesizing a Dyck State Graph with exceptional flow} \label{subsec: dcg-exn}

The Dyck State Graph (DSG) of a pushdown system is the subset of a pushdown system
reachable over legal paths.
(A path is \emph{legal} if it never tries to pop $a$ when a frame other than $a$ is on top of the stack.)
To synthesize a Dyck State Graph (DSG) from an 
(introspective) a pushdown system, 
Earl~\etal{} present an efficient, functional modification of the pushdown summarization algorithm~\cite{Earl:2012:IPDCFA}.
The algorithm iteratively constructs the reachable portion of the pushdown
transition relation by inserting $\epsilon$-summary edges whenever it finds empty-stack 
(\eg, push a, push b, pop b, pop a)
paths between control states.

For pushdown analysis  \emph{without exception handling}, only two kinds of transitions
can cause a change to the set of $\epsilon$-$\mathit{predecessors}$:
an intraprocedural empty-stack transition
and a frame-popping procedure return.
With the addition of $\mathbf{handle}$ frames to the stack, 
there are several new cases to consider for popping frames (and hence adding $\epsilon$-edges).

The following subsections highlight how to
handle exceptional flow during DSG synthesis, particularly as it relates
to maintaining $\epsilon$-summary edges.
The figures in these section use a graphical scheme
for describing the cases for $\epsilon$-edge insertion.
Existing edges are solid lines, while the $\epsilon$-summary edges 
to be added
are dotted lines.

\subsubsection{Intraprocedural push/pop of handle frames}

The simplest case is entering a {\tt try} block (a \textsf{push-handler}) 
and leaving a {\tt try} block (a \textsf{pop-handler}) entirely 
intraprocedurally---without throwing an exception.
Figure~\ref{fig:case1} shows such a case: if there is a handler push followed by a handler pop, 
the synthesized (dotted) edge must be added.

\subsubsection{Locally caught exceptions}
Figure~\ref{fig:case2} presents a case where 
a local handler catches an exception, popping it off the stack
and continuing.

\subsubsection{Exception propagation along the stack}

Figure~\ref{fig:case3} illustrates a case where 
an exception is not handled locally, and must pop off a 
call frame to reach the next handler on the stack.
In this case,
a popping self-edge from control state $q'$ to $q'$ 
lets the control state $q'$ see frames beneath the top.
Using popping self-edges, a single state can pop off as many frames as necessary 
to reach the handle---one at a time.

\subsubsection{Control transfers mixed in try/catch}

Figure~\ref{fig:case4} illustrates the situation where a procedure tries to
return while a $\mathbf{handle}$ frame is on the top of the stack.
It uses popping self-edges as well to find the top-most $\mathbf{call}$ frame.

\subsubsection{Uncaught exceptions}
The case in Figure~\ref{fig:case5} 
shows popping all frames back to  
the bottom of the stack---indicating an uncaught exception.

\begin{figure*}
\centering
\begin{minipage}{0.4\linewidth}
\begin{equation*}
  \xymatrix{ 
    \qstate_0 \ar[r]^{\exnframe_+} \ar @{.>} @(u,u) [rr]^\epsilon &
    \qstate' 
    \ar[r]^{\exnframe_-}
   & \qstate_1
  }  
\end{equation*}
\caption{Intraprocedural handler push/pop}
\label{fig:case1}
\end{minipage}%
\begin{minipage}{0.5\linewidth}
\begin{equation*}
  \xymatrix{ 
    \qstate_0 \ar[r]^{\callframe_+} &
    \qstate \ar[r]^{\exnframe_+}  \ar @{.>} @(u,u) [rrr]^\epsilon&
    \qstate' \ar[r]^{\epsilon} &
   \qstate'' \ar[r]^{\exnframe_-}
   & \qstate_1
     \ar[r]^{\epsilon} &
  }  
\end{equation*}
\caption{Locally caught exceptions}
\label{fig:case2}
\end{minipage}
\begin{minipage}{0.5\linewidth}
\begin{equation*}
  \xymatrix{ 
    \qstate_0 \ar[r]^{\exnframe_+}\ar[r]^{\exnframe_+} \ar @{.>} @(u,u) [rrr]^\epsilon &
    \qstate \ar[r]^{\callframe_+}  \ar @{.>} @(u,u) [r]^\epsilon&
    \qstate' \ar[r]^{\exnframe_-} \ar @{.>} @(dl,dr)_{\callframe_-} &
   \qstate_1
  } 
\end{equation*}
\caption{Exception propagation}
\label{fig:case3}
\end{minipage}%
\begin{minipage}{0.5\linewidth}
\begin{equation*}
  \xymatrix{ 
    \qstate_0 \ar[r]^{\callframe_+}  \ar @{.>} @(u,u) [rrr]^\epsilon &
    \qstate \ar[r]^{\exnframe_+}  \ar @{.>} @(u,u) [r]^\epsilon&
    \qstate' \ar @{.>} @(dl,dr)_{\exnframe_-} \ar[r]^{\callframe_-}&
    \qstate_1
  } 
\end{equation*}
\caption{Control transfers mixed in try/catch}
\label{fig:case4}
\end{minipage}
\begin{minipage}{0.5\linewidth}
\begin{equation*}
  \xymatrix{ 
    \qstate_0 \ar[r]^{\callframe_+}  &
    \qstate  \ar @{.>} @(dl,dr)_{\phrame^+_-}&
  } 
\end{equation*}
\caption{Uncaught exceptions}
\label{fig:case5}
\end{minipage}
\end{figure*}

\section{Evaluation} \label{sec:evaluation} \label{sec:eval}

We evaluated our pushdown exception flow analysis on standard Java
benchmarks from the DaCapo suite~\cite{local:DaCapo:paper} that we
were able to port to Android; we have also used some native Android
applications.
We ran these benchmarks on  
OS X 10.8.2 with a 64GB DDR3 memory,
2 Six-Core Intel Xeon X5675 CPUs, 3.07GHz machine.
Table~\ref{tbl:result} lists the results
for all applications.
To compare, we adopt metrics (and implementations) used by previous work~\cite{Fu:2005:rubust-java-server-apps,Bravenboer:2009:Exceptions}
for object-oriented programs: 
\begin{itemize}
\item VarPointsTo: Given a variable, to how many types may it point?
Smaller sets indicate higher data-flow precision.

\item ThrowPointsTo: At a throw, how many types of exceptions could be thrown?
Smaller sets indicate higher data-flow precision.

\item Exception-Catch-Link (E-C Link): A pair of instructions in which second catches the first.
Fewer E-C links indicate higher exception-flow precision.

\end{itemize}

The analysis result on running on Android applications of different size
have already demonstrated the promise of our analytic techniques, 
with the average one to three on VarPointsTo and ThrowPointsTo, and small 
number of E-C links.

The evaluation conducted on standard Java benchmarks
helps us compare results between our techniques and prior work.
We use the same version of benchmark suite, 
the DaCapo benchmark programs, v.2006-10.MR2,
which is used in~\cite{Bravenboer:2009:Exceptions}.
However, only {\tt{antlr, lucene, and pmd}}
run on Dalvik bytecode,
due to the Android SDK having class/interface naming clashes 
with the ones that are originally defined in Java SDK.

We contacted the authors for access to the original tool {\tt{Doop}}~\cite{Bravenboer:2009:Exceptions} 
to  run  the above benchmarks
and recompute the relative metrics.
Specifically, we ran {\tt{Doop}} Revision 958, on JRE 1.5 and Xubuntu 12.10
inside VirtualBox 4.2.2.
The metrics we compute are VarPointsTo, E-C Links and analysis run time.
with the option of context-sensitivity {\tt{1-Call+H}} and object-sensitivity{\tt{1-Obj+H}} 
respectively.
These options are the closest to the allocation strategy 
in our analysis: 1-call-site sensitivity for calls, and 1-object-sensitivity for object allocation.
In order to eliminate differences 
between the Dalvik and Java byte code,
the VarPointsTo metric computes 
how many types can be invoked on 
at each call site.

The comparison result is shown in the first three rows in Table~\ref{tbl:result}---the DaCapo benchmarks.
We could not get {\tt{Doop}} to operate properly on Android programs.

As we can see that
the pushdown exception-flow analysis produces 
almost two orders of magnitude improvement to the precision of points-to information 
and E-C Links for all three benchmarks over {\tt{Doop}}.
We have reported running times for completeness, but 
these numbers can't be compared as directly as precision.
\texttt{Doop} used a high-performance Datalog engine to solve
flow constraints; our implementation in Scala is asymptotically efficient,
but it is not optimized; it incurs a significant constant-factor overhead.

The effect of analysis time varies from different benchmarks.
But take into consideration of the difference of running environment,
{\tt{Doop}} demonstrated less analysis than our analysis does.
However, the co-analysis of pushdown system and augmented abstract garbage collection 
has demonstrated the best precision/performance trade-offs.

In Table~\ref{tbl:result}, 
adding garbage collection and live-range analysis restriction ({\tt{+gc+lra}}) 
improves analysis time  more significantly for Android application than Java applications.
The reason is that Android applications are more sensitive to the LRA due to Android's multi-entry points structure.
However, the results on the DaCapo benchmarks clearly indicate improvements over \texttt{Doop} in precision.

\begin{table}
{\footnotesize
    \begin{tabular}{| c| c | c| c | c | c | c | c | c |}
    \hline
    \textbf{Benchmark} & \textbf{LOC} & \textbf{Opts} & \textbf{Nodes} & \textbf{Edges} &  \textbf{VarPointsTo} & \textbf{Throws} & \textbf{E-Cs} & \textbf{Time} \\ \hline 
   antlr &   \multirow{4}{*}{35000} &pdcfa & $\infty$  & $\infty$   & $\infty$  &  $\infty$   &  $\infty$  & $\infty$  
     \\ & & +gc+lra & 1212  & 1251  &   (681, 2)& (78,2) &  65 & 4135 \\
      & & 1-Call+H & -  & -  &   (40503, 614)& - &  2277 & \textgreater4h \\
      & & 1-Obj+H & -  & -  &   (41339,626) & - &  2203 & \textgreater3h \\
     \hline
         lusearch &  \multirow{4}{*}{ 87000}& pdcfa & 91574  & 105154  &  (916, 3) &  (309, 3)  & 76 & 5520 
          \\ && +gc+lra & 14646  & 16045  &   (709, 2)& (213,2) & 59 & 2785 \\
            & & 1-Call+H & -  & -  &   (22970, 348)& - &  2378 &  2796 \\
                & & 1-Obj+H & -  & -  &   (24225,367) & - &  2304 & 1548 \\
          \hline
         pmd & \multirow{4}{*}{55000} &pdcfa & 59173  & 61162  &  (1432, 3) &  (103, 3)  & 51  & 4351
         \\ && +gc+lra & 3537  & 4035  &   (1017, 2)& (61,2) & 38& 1323 \\
          & & 1-Call+H & -  & -  &   (25286, 383)& - &  2284 &  3375 \\
          & & 1-Obj+H & -  & -  &   (26049,395) & - &  2212 & 3413 \\
         \hline
          \hline
                    Butane &  \multirow{2}{*}{2506}&pdcfa & 20140  & 20334 &  (463,3) &  (2,1)  &  2 &   920
                     \\ &&+gc+lra & 724  & 740  &   (322, 2)& (2,1) & 2 & 676 \\
                     \hline
       UltraCoolMap & \multirow{2}{*}{2605} &pdcfa & 17044  & 17047  &  (552, 2) &  (2, 1)  & 2  & 1056 
                                \\ &&+gc+lra & 948  & 951  &   (465, 2)& (2,1) & 2 & 28 \\
                                \hline
        
               Sysmon & \multirow{2}{*}{4429}& pdcfa &  $\infty$ & $\infty$ & $\infty$ &  $\infty$  & $\infty$  & $\infty$
                                               \\ && +gc+lra & 3191  & 3431  &   (864, 2)& (10,1) &10& 1534 \\ \hline
               TodoList & \multirow{2}{*}{6092}& pdcfa & 6020  & 6406  &  (387,2) & (0,0)  & 0  & 1766
                \\ && +gc+lra & 1348  & 1434  &   (280, 1)& (0,0) & 0& 224 \\
                \hline
                   SwiFTP & \multirow{2}{*}{6521}& pdcfa & 147785  & 155656  &  (634,2) &  (3, 1)  &  3 & 4277
                                                         \\ && +gc+lra & 4711  & 5053  &   (564, 2)& (1,1) & 1 & 1871 \\
                                                         \hline
                  MediaFun&  \multirow{2}{*}{7815}& pdcfa & $\infty$  & $\infty$&  $\infty$&  $\infty$ &  $\infty$ &  $\infty$
                                    \\ && +gc+lra & 9894  & 10298  &   (674, 2)& (13,1) & 10 & 3032 \\
                                    \hline                                      
               AndroidGame  &\multirow{2}{*}{63755} &pdcfa & 3831  & 4001  &  (405, 3) &  (2, 1)  &  2 & 989 
                                                                    \\  && +gc+lra & 1111  & 1180  &   (363, 2)& (2,1) & 2& 246 \\
                                                                    \hline
                                                         
                     ConnectBot &\multirow{2}{*}{68382}& pdcfa & 14484  & 15147  &  (543, 3) &  (5, 1)  & 5  & 3012 
                     \\ && +gc+lra & 3739  & 4063  &   (368, 1)& (5,1) & 5& 1896 \\
                     \hline
                                     
     \hline
    \end{tabular}    
    }
    \caption{Benchmark results: {\tt{VarPointsTo}} and {\tt{Throws}} is 
    presented as tuples $(a,b)$, where $a$ is the total entries, $b$ is the average 
    types being invoked on in {\tt{VarPointsTo}} case,
     and average exception objects thrown in {\tt{Throws}} case. 
     All times are in seconds.
     $\infty$ denotes the analysis did not finish within 6000 seconds.}
     \label{tbl:result}
\end{table}

\section{Related Work} \label{sec: related}

Precise and scalable context-sensitive points-to analysis 
has been an open problem for decades.
Progress in general has been gradual,
with results like object-sensitivity~\cite{local:Milanova:2007:LCP,local:Milanova:2005:parameterizedobject} intermittently providing a leap for most programs.
Most results target improvements for individual classes of programs.
The techniques we present here broadly target at all programs, and it is orthogonal to and compatible with
results like object-sensitivity.

Much work in pointer analysis exploits methods to improve performance
by strategically reducing precision. 
Lattener~\etal~show that an analysis with a context-sensitive
heap abstraction can be efficient by sacrificing precision 
under unification constraints~\cite{local:Lattner:2007:MCP}.

In full-context-sensitive pointer analysis, the literature has sought
 context abstractions that provide precise pointer information
while not sacrificing performance.
Milanova found that an object-sensitive analysis~\cite{local:Milanova:2005:parameterizedobject}
 is an effective context abstraction for object-oriented programs. 
 This is confirmed by the extensive evaluation by Lhotak~\cite{local:Lhotak:2008:EBC}.
He and other researchers have also argued for using context-sensitive heap abstraction 
to improve precision\cite{local:Nystrom:2004:IHS}.

BDDs have been used to compactly represent the large  amount of redundant  data in context-sensitive pointer analysis 
efficiently~\cite{local:Berndl:2003:PAU,Whaley:2004:CCP,local:Xu:2008:MEC}. 
Specifically, Xu and Routev's work~\cite{local:Xu:2008:MEC} reduces the redundancy by choosing the right context abstractions.
Such advancements could be applied to our pushdown framework, as they are orthogonal to its central thesis.

\paragraph{\textbf{Finite-state analysis of exceptions}}

The main contribution of the paper is significantly improved analysis precision
via pushdown systems that analyze the exceptional control-flow of object-oriented programs.

The bulk of the previous literature has focused on finite-state abstractions for Java programs, \ie,
$k$-CFA and its variants.
Specifically, for the work that handles exception flows, the analysis is based on 
context-insensitivity or a limited form of context-sensitivity, which makes them unable to differentiate the contexts 
of where an exception is thrown  and what handlers precisely can handle the exception.
Robillard~\etal~\cite{Robillard:2003:evolution-exception} 
presents a truly interprocedural exception-flow analysis, but exceptions propagate
via imprecise control flows by using class hierarchy analysis.
The same is true for Jo~\etal{}~\cite{local:Jo:2002:UncaughtException},
and its extension for concurrent Java programs~\cite{Ryu:2001:MultiThreadExceptions}.
Leroy and Pessaux~\cite{Leroy:2004:TypeBasedUncaughtExceptions} use
type systems to model exceptions, specifically to analyze uncaught
exceptions. Limited context-sensitivity is employed for the purpose of
more precise results on polymorphic functions.
 Fu~\etal{}~\cite{Fu:2005:rubust-java-server-apps} proposed the E-C link metric to evaluate 
 exception-flow precision. They also documented the exception handler matching problem caused 
 by an imprecise control flow graph. They approach the problem by employing  points-to information to refine
 control-flow reachability.
 Bravendoer and Smaragdakis ~\cite{Bravenboer:2009:Exceptions}
 propose to join points-to analysis and exception flow analysis to improve 
 precision and analysis run time in their Doop framework, 
 based on the optimized analysis engine using Datalog~\cite{Bravebboer:2009:declare-pointsto}.
 They have conducted extensive comparison of different options for polyvariance.
 It is the most precise and efficient exception-flow analysis compared to other work, with respect of points-to 
and E-C links. We conduct our comparison with respect to their work, and found the 
pushdown approach can yield significant improvement in precision, but the run-time is not comparable 
to their work, partly due to their mature optimization methodology for Datalog.

\paragraph{\textbf{Pushdown analysis for the $\lambda$-calculus}}

Vardoulakis and Shivers's CFA2~\cite{mattmight:Vardoulakis:2010:CFA2}
is the precursor to the pushdown control-flow
analysis~\cite{local:Earl:2010:PDCFA}.
CFA2 is a table-driven summarization algorithm that exploits the
balanced nature of calls and returns to improve return-flow precision
in a control-flow analysis.
While CFA2 uses a concept called ``summarization,'' it is a
summarization of execution paths of the analysis, roughly equivalent
to Dyck state graphs.

In terms of recovering precision, pushdown control-flow
analysis~\cite{local:Earl:2010:PDCFA} is the dual to abstract garbage
collection:
it focuses on the global interactions of configurations via
transitions to precisely match push-pop/call-return, thereby
eliminating all return-flow merging.
However, pushdown control-flow analysis does nothing to improve
argument merging.

This work directly draws on our previous work 
on pushdown analysis for higher-order programs~\cite{local:Earl:2010:PDCFA} and
 introspective pushdown system (IPDS) for higher-order programs~\cite{Earl:2012:IPDCFA}.
IPDS  has tackled the challenge of incorporating abstract garbage collection~\cite{mattmight:Might:2006:GammaCFA}
into pushdown system and improving the summarization algorithm  for efficiency.
That work shows significant  improvements in precision and analysis time for the $\lambda$-calculus.
We extend the introspective work in two dimensions: (1)  we generalize the framework (including abstract garbage collection) to an object-oriented language,
and (2) we adapt the Dyck state graph synthesis algorithm to handle the new stack change behavior
introduced by exceptions.

\paragraph{\textbf{CFL- and pushdown-reachability techniques}}
In previous work, Earl~\etal~\cite{Earl:2012:IPDCFA}
develop a pushdown reachability algorithm
suitable for the pushdown systems that we generate.
It essentially draws on CFL- and pushdown-reachability
analysis~\cite{mattmight:Bouajjani:1997:PDA-Reachability,mattmight:Kodumal:2004:CFL,mattmight:Reps:1998:CFL,mattmight:Reps:2005:Weighted-PDA}.
For instance, \ecg s, or equivalent variants thereof, appear in many
context-free-language and pushdown reachability algorithms.
Dyck state graph synthesis is an attractive perspective
on pushdown reachability because it is purely functional,
and it allows targeted modifications to the algorithm.

CFL-reachability techniques have also been used to compute classical
finite-state abstraction CFAs~\cite{mattmight:Melski:2000:CFL} and
type-based polymorphic control-flow
analysis~\cite{mattmight:Rehof:2001:TypeBased}.
These analyses should not be confused with pushdown control-flow
analysis, which is computing a fundamentally different kind of CFA.

\paragraph{\textbf{Pushdown exception-flow analysis}}
There is little work on pushdown analysis for object-oriented langages as a whole.
Sridharan and Bodik proposed demand-driven analysis for Java that
matches reads with writes to object fields selectively, by using
refinement~\cite{Manu:2006:RefinementJava}.  They employ a
refinement-based CFL-reachability technique that refines calls and
returns to valid matching pairs, but approximates for recursive calls.
They do not consider specific applications of CFL-reachability to
exception-flow.

\section{Conclusion} \label{sec:conclusion}

Poor analysis of exceptions pollutes the interprocedural control-flow analysis
of a program.
In order to model exceptional control-flow precisely, we abandoned traditional
finite-state approaches (e.g. $k$-CFA and its variants).
In its place, we generalized pushdown control-flow analysis from the
$\lambda$-calculus~\cite{Earl:2012:IPDCFA} to object-oriented programs, and
made it capable of handling exceptions in the process.
Pushdown control-flow analysis models the program stack (precisely) with the
pushdown stack.
Computing the reachable control states of the pushdown system (its Dyck state
graph) yields combined data- and control-flow analysis of a program.
Comparing this approach to the
state-of-the-art~\cite{mattmight:Bravenboer:2009:Exceptions}, shows
substantially improved precision.
To improve time, we adapted abstract garbage collection to object-oriented
program analysis.
The end result is an improvement in data- and control-flow precision of roughly
two orders of magnitude when soundly reasoning  in the presence of exceptions.

\bibliographystyle{acm}
\bibliography{mattmight,shuyingliang,local}

\vfill
\pagebreak

\appendix

\parindent=0in
\parskip=0.1in

\section{Optional Appendix}

This appendix is not required reading.
We provide it for background, refreshment and deeper discussion.

\subsection{Notational conventions}

We strive to stick to conventional notation
and names wherever possible.
For less common concepts,
we define them here.

The functional update operation $f[x \mapsto y]$
extends a function with a new binding:
\begin{align*}
f[x \mapsto y](x) &= y
\\
f[x \mapsto y](x') = f(x')&~\text{if} ~x \ne x'
\text.
\end{align*}

For objects like stores, we lift 
the least upper bound $\join$ point-wise:
\begin{align*}
 (\astore \join \astore')(\aaddr) = \store(\aaddr) \union \store'(\aaddr)
 \text.
\end{align*}

\subsection{Additional concrete transition relations} \label{appendix:simple-conc-rules}
\subsubsection{Stepping over nops and labels}: 
The simplest instruction nop does not change any component in the configuration state:
 \begin{align*}
  \overbrace{(\sembr{\syn{nop}} :  \vec{s}, \fp, \store, \cont)}^{\conf}
  &\To
  \overbrace{(\vec{s},\fp,\store, \cont)}^{\conf'} 
 \text.
  \end{align*}
$\syn{label}$ and $\syn{line}$ statement shares the same transition form.
 \footnote{The $\syn{line}$ statement is mainly for instrumenting \textit{context} information to the statements that are actually interpreted.}

\subsubsection{Unconditional jumps}: 
This kind of statement forces program to jump to the target statement sequence:
 \begin{align*}
  \overbrace{(\sembr{(\syn{goto} ~\mathit{label})} :  \vec{s}, \fp, \store, \cont)}^{\conf}
  &\To
  \overbrace{(S(\mathit{label}),\fp,\store, \cont)}^{\conf'} 
 \text.
  \end{align*}
where the function $\mathit{S} : \syn{Label} \rightarrow \syn{Stmt^*}$ 
maps a label to the statement sequence starting with that label.

\subsubsection{Conditionals}
The if-goto is not much more complicated than a nop or goto, but it needs to evaluate the conditional expression:
 \begin{align*}
  \overbrace{(\sembr{(\syn{if} ~\ae~(\syn{goto}~\mathit{label}))} :  \vec{s}, \fp, \store, \cont)}^{\conf}
  &\To
  \overbrace{
   \begin{cases}
      (\vec{s}, \fp, \store, \cont)
      &  {\ArgEval(\ae, \fp,\store) \ne \mathit{false}}
      \\
       (S(\mathit{label}),\fp,\store, \cont) 
      &  \text{otherwise} 
      \end{cases}}^{\conf'}
 \text.
  \end{align*}
  
\subsubsection{Atomic assignments}

Atomic assignment statements assign the value of an atomic expression to a variable(register).
This involves evaluating the expression, calculating the frame address 
to modify and then updating the store.

\begin{align*}
  \overbrace{(\syn{assgin}~\mathit{\$name} ~\expr)}^{\conf}
  &\To
  \overbrace{(\vec{s},\fp,\store', \cont)}^{\conf'} 
 \text{where}
 \\
 \store' &= \store[(\fp, \mathit{\$name}) \mapsto \ArgEval(e, \fp, \store)] 
  \end{align*}
Note that a large set of instruction statements are transformed 
into $\syn{assign}$ form.
For example, ($\syn{move}\mhyphen\syn{result} ~\$name) $ is transformed to ~$(\syn{assign} ~\$name~ \syn{ret})$ form.

\subsection{Other abstract transition relations} \label{appendix-simple-abs-trs}

\begin{itemize}
\item{\textbf{Stepping over nops and labels}}: 
Abstract transition relations for this kind of statement is almost like correspondent concrete semantics (Section~\ref{subsec:conc-tr})
 \begin{align*}
  \overbrace{(\sembr{\syn{nop}} :  \vec{s}, \afp, \astore, \acont)}^{\aconf}
  &\aTo
  \overbrace{(\vec{s},\afp,\astore, \acont)}^{\aconf'} 
 \text.
  \end{align*}
 The same right hand side of transition relation for $\syn{label}$ and $\syn{line}$ statement.
 
\item{\textbf{Unconditional jumps}}: 
Like nop statement and label statements, no much difference but with abstract components replaced.
 \begin{align*}
  \overbrace{(\sembr{(\syn{goto} ~\mathit{label})} :  \vec{s}, \afp, \astore, \acont)}^{\aconf}
  &\aTo
  \overbrace{(S(\mathit{label}),\afp,\astore, \acont)}^{\aconf'} 
 \text.
  \end{align*}

\item{\textbf{Conditionals}}
The if-goto is not much more complicated than a nop or goto, but it needs to evaluate the conditional expression:
 \begin{align*}
  \overbrace{(\sembr{(\syn{if} ~\ae~(\syn{goto}~\mathit{label}))} :  \vec{s}, \afp, \astore, \acont)}^{\aconf}
  &\aTo
  \overbrace{
   \begin{cases}
      (\vec{s}, \afp, \astore, \acont)
      &  {\mathit{false} \notin \aArgEval(\ae, \afp,\astore)}
      \\
       (S(\mathit{label}),\afp,\astore, \acont) 
      & \text{otherwise}
      \end{cases}}^{\aconf'}
 \text.
  \end{align*}
  
\item{\textbf{Atomic assignments}}: 
The main change to the abstract transition relation for atomic assignments 
resides in the operation to the store component:
\begin{align*}
  \overbrace{(\syn{assgin}~\mathit{\$name} ~\expr)}^{\aconf}
  &\To
  \overbrace{(\vec{s},\afp,\astore', \acont)}^{\aconf'} 
 \text{where}
 \\
 \astore' &= \astore \join [(\afp, \mathit{\$name}) \mapsto \aArgEval(e, \afp, \astore)] 
  \end{align*}

\end{itemize}
\subsection{Syntactic sugar for pushdown systems}

When a triple $(x,\ell,x')$ is an edge in a labeled graph:
\begin{equation*}
x \pdedge^\ell x'  \equiv
(x,\ell,x')
\text.
\end{equation*}
Similarly, when a pair $(x,x')$ is a graph edge:
\begin{equation*}
\biedge{x}{x'} \equiv (x,x')
\text.
\end{equation*}
We use both
string and vector notation for sequences:
\begin{equation*}
a_1 a_2 \ldots a_n \equiv \vect{a_1,a_2,\ldots,a_n}
\equiv
\vec{a} 
\text.
\end{equation*}

\subsection{Stack actions, stack change and stack manipulation}

Stacks are sequences over a stack alphabet $\StackAlpha$.
To reason about stack manipulation concisely,
we first turn stack alphabets into ``stack-action'' sets;
each character represents a change to the stack: push, pop or no
change.

For each character $\stackchar$ in a stack alphabet $\StackAlpha$, the
\defterm{stack-action} set $\StackAlpha_\pm$ contains a push character
$\stackchar_{+}$; a pop character $\stackchar_{-}$;
and a no-stack-change indicator, $\epsilon$:
\begin{align*}
\stackact \in \StackAlpha_\pm &\produces \epsilon && \text{[stack unchanged]} 
\\
    &\;\;\opor\;\; \stackchar_{+}  \;\;\;\text{ for each } \stackchar \in \StackAlpha && \text{[pushed $\stackchar$]}
    \\
      &\;\;\opor\;\; \stackchar_{-}  \;\;\;\text{ for each } \stackchar \in \StackAlpha && \text{[popped $\stackchar$]}
      \text.
      \end{align*}
      In this paper, the symbol $\stackact$ represents some stack action.

      When we develop introspective pushdown systems, we are going
      to need formalisms for easily manipulating stack-action strings
      and stacks.
      Given a string of stack actions, we can compact it into a minimal
      string describing net stack change.
      We do so through the operator $\fnet{\cdot} : \StackAlpha_\pm^* \to
      \StackAlpha_\pm^*$, which cancels out opposing adjacent push-pop stack
      actions:
      \begin{align*}
      \fnet{\vec{\stackact} \; \stackchar_+\stackchar_- \; \vecp{\stackact}} &= 
      \fnet{\vec{\stackact} \; \vecp{\stackact}} 
      &
      \fnet{\vec{\stackact} \; \epsilon \; \vecp{\stackact}} &= 
      \fnet{\vec{\stackact} \; \vecp{\stackact}} 
      \text,
      \end{align*}
      so that
      $\fnet{\vec{\stackact}} = \vec{\stackact}\text,$
      if there are no cancellations to be made in the string $\vec{\stackact}$.

      We can convert a net string back into a stack by stripping off the
      push symbols with the stackify operator, $\fstackify{\cdot} :
      \StackAlpha^{*}_\pm \parto \StackAlpha^*$:
      \begin{align*}
      \fstackify{\stackchar_+ \stackchar_+' \ldots \stackchar_+^{(n)}} =
      \vect{\stackchar^{(n)}, \ldots, \stackchar', \stackchar}
      \text,
      \end{align*}
      and for convenience, $[\vec{\stackact}] = 
      \fstackify{\fnet{\vec{\stackact}}}$.
      Notice the stackify operator is defined for strings containing
      only push actions. 

      \subsection{Pushdown systems}
        A \defterm{pushdown system} is a triple
        $M = (\ControlStates,\StackAlpha,\transfunction)$ where:
        \begin{enumerate}

        \item $\ControlStates$ is a finite set of control states;

        \item $\StackAlpha$ is a stack alphabet; and

        \item $\transfunction \subseteq
        \ControlStates \times \StackAlpha_\pm \times \ControlStates$ is a transition relation.
        \end{enumerate}
        The set $\ControlStates \times \StackAlpha^*$ is 
        called the \defterm{configuration-space} of this pushdown system.
        We use $\mathbb{PDS}$ to denote the class of all pushdown systems.
        \\

        \noindent
        For the following definitions, let $M = (\ControlStates,\StackAlpha,\transfunction)$.
        \begin{itemize}

        \item The labeled \defterm{transition relation} $(\PDTrans_{M}) \subseteq
        (\ControlStates \times \StackAlpha^*) \times 
        \StackAlpha_\pm \times 
        (\ControlStates \times \StackAlpha^*)$
        determines whether one configuration may transition to another while performing the given stack action:
        \begin{align*}
(\qstate, \vec{\stackchar}) 
  \mathrel{\PDTrans_M^\epsilon}
  (\qstate',\vec{\stackchar}) 
  & \text{ iff }
  \qstate \pdedge^\epsilon \qstate'
  \in \transfunction
  && \text{[no change]}
  \\
    (\qstate, \stackchar : \vec{\stackchar}) 
    \mathrel{\PDTrans_M^{\stackchar_{-}}}
    (\qstate',\vec{\stackchar})
    & \text{ iff }
    \qstate \pdedge^{\stackchar_{-}} \qstate'
    \in \transfunction
    && \text{[pop]}
    \\
      (\qstate, \vec{\stackchar}) 
      \mathrel{\PDTrans_{M}^{\stackchar_{+}}}
      (\qstate',\stackchar : \vec{\stackchar}) 
      & \text{ iff }
      \qstate \pdedge^{\stackchar_{+}} \qstate'
      \in \transfunction
      && \text{[push]}
      \text.
      \end{align*}

      \item If unlabelled, the transition relation $(\PDTrans)$ checks whether \emph{any} stack action can enable the transition:
      \begin{align*}
      \conf \mathrel{\PDTrans_{M}} \conf' \text{ iff }
      \conf \mathrel{\PDTrans_{M}^{\stackact}} \conf' \text{ for some stack action } \stackact 
      \text.
      \end{align*}

      \item

      For a string of stack actions $\stackact_1 \ldots
      \stackact_n$:
      \begin{equation*}
      \conf_0 \mathrel{\PDTrans_M^{\stackact_1\ldots\stackact_n}} \conf_n
      \text{ iff }
      \conf_0
      \mathrel{\PDTrans_M^{\stackact_1}} 
      \conf_1
      \mathrel{\PDTrans_M^{\stackact_2}} 
      \cdots
      \mathrel{\PDTrans_M^{\stackact_{n-1}}} 
      \conf_{n-1} 
      \mathrel{\PDTrans_M^{\stackact_n}} 
      \conf_n\text,
      \end{equation*}
      for some configurations $\conf_0,\ldots,\conf_n$.

      \item

      For the transitive closure:
      \begin{equation*}
      \conf \mathrel{\PDTrans_M^{*} \conf'
        \text{ iff }
        \conf \mathrel{\PDTrans_M^{\vec{\stackact}}} \conf'
          \text{ for some action string }
        \vec{\stackact}} 
        \text.
        \end{equation*}

        \end{itemize}
      \subsection{Rooted pushdown systems}

        A \defterm{rooted pushdown system} is a quadruple 
        $(\ControlStates,\StackAlpha,\transfunction,\qstate_0)$ in which 
        $(\ControlStates,\StackAlpha,\transfunction)$ is a pushdown system and
        $\qstate_0 \in \ControlStates$ is an initial (root) state.
        $\mathbb{RPDS}$  is the class of all rooted pushdown
        systems.

        For a rooted pushdown system $M =
        (\ControlStates,\StackAlpha,\transfunction,\qstate_0)$, we define 
        the \defterm{reachable-from-root transition relation}:
        \begin{equation*}
        \conf \RPDTrans_M^{\stackact} \conf' \text{ iff } 
(\qstate_0,\vect{})
  \mathrel{\PDTrans_M^*} 
  \conf 
  \text{ and }
  \conf 
  \mathrel{\PDTrans_M^\stackact}
  \conf'
  \text.
  \end{equation*}
  In other words, the root-reachable transition relation also makes
  sure that the root control state can actually reach the transition.

  We overload the root-reachable transition relation to operate on
  control states:
  \begin{equation*}
  \qstate 
  \mathrel{\RPDTrans_M^\stackact}
  \qstate' \text{ iff }
(\qstate,\vec{\stackchar}) 
  \mathrel{\RPDTrans_M^\stackact}
  (\qstate',\vecp{\stackchar}) 
  \text{ for some stacks }
  \vec{\stackchar},
  \vecp{\stackchar}
  \text.
  \end{equation*}
  For both root-reachable relations, if we elide the stack-action label,
  then, as in the un-rooted case, the transition holds if \emph{there
    exists} some stack action that enables the transition:
    \begin{align*}
    \qstate 
    \mathrel{\RPDTrans_M}
    \qstate' \text{ iff }
    \qstate 
    \mathrel{\RPDTrans_M^{\stackact}}
    \qstate' 
    \text{ for some action } \stackact
    \text.
    \end{align*}

    \subsection{Computing reachability in pushdown systems}

    A pushdown flow analysis 
    can be construed as
    computing the \emph{root-reachable}
    subset of control states in a rooted pushdown system, 
    $M = (\ControlStates,\StackAlpha,\transfunction,\qstate_0)$:
    \begin{align*}
    \setbuild{\qstate}{
      \qstate_0 
        \mathrel{\RPDTrans_M}
      \qstate
    }
\end{align*}
Reps~\emph{et. al} and many others provide a straightforward ``summarization'' algorithm
to compute this set~\cite{mattmight:Bouajjani:1997:PDA-Reachability,mattmight:Kodumal:2004:CFL,mattmight:Reps:1998:CFL,mattmight:Reps:2005:Weighted-PDA}.
Our preliminary report also offers a reachability algorithm 
tailored to higher-order
programs~\cite{local:Earl:2010:PDCFA}.

\subsection{Nondeterministic finite automata}
In this work, we will need a finite description of 
all possible stacks at a given control state within
a rooted pushdown system.
We will exploit the fact that the set of stacks
at a given control point is a regular language.
Specifically, we will extract a nondeterministic finite automaton
accepting that language from the structure
of a rooted pushdown system.
A \defterm{nondeterministic finite automaton} (NFA) is a quintuple
$M = (\QStates, \Alphabet, \transfunction, \qstate_0, \FStates)$:
\begin{itemize}
\item $\ControlStates$ is a finite set of control states;

\item $\Alphabet$ is an input alphabet; 

\item $\transfunction 
  \subseteq
\ControlStates \times (\Alphabet \union \set{\epsilon}) 
  \times \ControlStates$ 
  is a transition relation.

  \item $\qstate_0$ is a distinguished start state.

  \item $\FStates \subseteq \QStates$ is a set of accepting states.
  \end{itemize}
  We denote the class of all NFAs as $\mathbb{NFA}$.

   \subsection{Introspective pushdown systems}

    An \defterm{introspective pushdown system} is a quadruple
    $M = (\ControlStates,\StackAlpha,\transfunction,\qstate_0)$:
    \begin{enumerate}
  
    \item $\ControlStates$ is a finite set of control states;
  
    \item $\StackAlpha$ is a stack alphabet; 
  
    \item $\transfunction \subseteq \ControlStates \times \StackAlpha^*
    \times \StackAlpha_\pm \times \ControlStates$ is a transition relation; and
  
    \item $\qstate_0$ is a distinguished root control state. 
    \end{enumerate}
    The second component in the transition relation is 
    a realizable stack at the given control-state.
    This realizable stack distinguishes an introspective pushdown system
    from a general pushdown system.
    $\mathbb{IPDS}$  denotes the class of all introspective pushdown
    systems.

    Determining how (or if) a control state $\qstate$
    transitions to a control state $\qstate'$, requires knowing a
    path taken to the state $\qstate$.
    Thus, we need to define reachability inductively.
    When $M = (\ControlStates,\StackAlpha,\transfunction,\qstate_0)$,
    transition from the initial control state considers only
    empty stacks:
    \begin{align*}
    \qstate_0  
    \mathrel{\RPDTrans_M^\stackact}
    \qstate
    \text{ iff }
  (\qstate_0, \vect{}, \stackact, \qstate)
    \in \transfunction
    \text.
    \end{align*}
    For non-root states, the paths to that state matter,
    since they determine the stacks realizable with that state:
    \begin{align*}
    \qstate 
    \mathrel{\RPDTrans_M^\stackact}
    \qstate' 
    &\text { iff there exists }
    \vec{\stackact}
    \text{ such that }
    \qstate_0
    \mathrel{\RPDTrans_M^{\vec{\stackact}}}
    \qstate
    \text{ and }
    (\qstate, [\vec{\stackact}], \stackact, \qstate' )
    \in
    \transfunction,
    \\
      &  \text{ where }\qstate
      \mathrel{\RPDTrans_M^{\vect{\stackact_1,\ldots,\stackact_n}}}
      \qstate'
      \text{ iff }
      \qstate
      \mathrel{\RPDTrans_M^{\stackact_1}}
      \qstate_1
      \mathrel{\RPDTrans_M^{\stackact_2}}
      \cdots
      \mathrel{\RPDTrans_M^{\stackact_n}}
      \qstate'
      \text.
      \end{align*}
      
      \subsection{Computing reachability in intropective pushdown system}
      We cast our reachability algorithm for
          introspective pushdown systems 
          as
          finding a fixed-point, in
          which we incrementally accrete the reachable control states
          into a ``Dyck state graph.''
      
          A \defterm{Dyck state graph} is a quadruple
          $G = (\DSStates,\DSFrames,\DSEdges,\dsstate_0)$, in which:
          \begin{enumerate}
          \item $\DSStates$ is a finite set of nodes;
          \item $\DSFrames$ is a set of frames;
          \item $\DSEdges \subseteq \DSStates \times \DSFrames_\pm \times
          \DSStates$ is a set of stack-action edges; and
          \item $\dsstate_0$ is an initial state;
          \end{enumerate}
          such that for any node $\dsstate \in \DSStates$, it must be the case that:
          \begin{equation*}
      (\dsstate_0,\vect{}) \mathrel{\PDTrans_G^*} (\dsstate,\vec{\dsframe})
        \text{ for some stack } \vec{\dsframe}
        \text.
        \end{equation*}
        In other words, a Dyck state graph is equivalent to a rooted
        pushdown system in which there is a legal path to every control state
        from the initial control state.\footnote{We chose the term \emph{Dyck
          state graph} because the sequences of stack actions along valid paths
            through the graph correspond to substrings in Dyck languages.
            A \defterm{Dyck language} is a language of balanced, ``colored''
            parentheses.
            In this case, each character in the stack alphabet is a color.}
            We use $\mathbb{DSG}$ to denote the class of Dyck state graphs.
      (Clearly, $\mathbb{DSG} \subset \mathbb{RPDS}$.)

        Our goal is to compile an 
        implicitly-defined introspective pushdown system into 
        an explicited-constructed Dyck state graph.
        During this transformation, the per-state path
        considerations
        of an introspective pushdown are ``baked into''
        the Dyck state graph.
        We can formalize this compilation process as a map, 
        $\fDSG : \mathbb{IPDS} \to
        \mathbb{DSG}\text.$
      
        Given an introspective pushdown system $M =
        (\ControlStates,\StackAlpha,\transfunction,\qstate_0)$, its equivalent Dyck state
        graph is $\fDSG(M) = (\DSStates,\DSFrames,\DSEdges,\qstate_0)$, where 
        $\dsstate_0 = \qstate_0$,
        the set $\DSStates$ contains reachable nodes:
        \begin{equation*}
        \DSStates = 
        \setbuild{ \qstate }{ 
            \qstate_0
            \mathrel{\RPDTrans_M^{\vec{\stackact}}}
            \qstate
            \text{ for some stack-action sequence }
            \vec{\stackact}
            }
      \text,
        \end{equation*}
        and the set $\DSEdges$ contains reachable edges:
        \begin{equation*}
        \DSEdges = \setbuild{ \qstate \pdedge^{\stackact} \qstate' }{
            \qstate
            \mathrel{\RPDTrans_M^{\stackact}}
          \qstate'
        }
      \text.
      \end{equation*}
      Our goal is to find a method
      for computing a Dyck state graph from 
      an introspective pushdown
      system.
      
          \subsection{Garbage collection in introspective pushdown systems } \label{subsubsec: gc-ipds}
          Having augmented the abstract garbage collection with respect of objects, 
         we are now ready to embed it into introspective pushdown systems.
         using the function $\afIPDS : \syn{Stmt^*} \to \mathbb{IPDS}$ as presented in  
          Fig \ref{fig:acesk-to-ipds} .

          \begin{figure*}
          \centering
          \begin{minipage}[!htbp]{0.55\linewidth}
             \begin{align*}
         \afPDS(\expr) &= (\QStates,\StackAlpha,\transfunction,\qstate_0)
           \text{, where }
           \\
             \QStates &= \syn{Stmt*} \times \sa{FramePointer} \times \sa{Store}
             \\
               \StackAlpha &= \sa{Frame}
               \\
                 (\qstate,\epsilon,\qstate') \in \transfunction
                 & \text{ iff }
         (\qstate, \acont)
           \aTo
           (\qstate', \acont)
           \text{ for all } \acont
           \\
             (\qstate,\aphrame_{-},\qstate') \in \transfunction
             & \text{ iff }
         (\qstate, \aphrame : \acont)
           \aTo
           (\qstate',\acont)
           \text{ for all } \acont
           \\
             (\qstate,{\aphrame_+},\qstate') \in \transfunction
             & \text{ iff }
         (\qstate, \acont)
           \aTo
           (\qstate', \aphrame: \acont)
           \text{ for all } \acont
             \text.
             \end{align*}
             \caption{$\afPDS : \syn{Exp} \to \mathbb{RPDS}$.
             }
         \label{fig:acesk-to-pds}
          \end{minipage}%
          \begin{minipage}[!htbp]{0.6\linewidth}
             \begin{align*}
          \afIPDS(\expr) &= (\QStates,\StackAlpha,\transfunction,\qstate_0)
            \\
              \QStates &= \syn{Stmt^*} \times \sa{FramePointer} \times \sa{Store}
              \\
                \StackAlpha &= \sa{Frame}
                \\
                  (\qstate,\acont,\epsilon,\qstate') 
                  &\in \transfunction
                   \text{ iff }
          \aCollect(\qstate, \acont)
            \aTo
            (\qstate', \acont)
            \\
              (\qstate,\aphrame : \acont,\aphrame_{-},\qstate')
              &\in \transfunction
               \text{ iff }
          \aCollect(\qstate, \aphrame : \acont) 
            \aTo
            (\qstate',\acont)
            \\
              (\qstate,\acont,\aphrame_{+},\qstate') 
              &\in \transfunction
               \text{ iff }
          \aCollect(\qstate, \acont)
            \aTo
            (\qstate',\aphrame : \acont)
              \text.
              \end{align*}
                     \caption{ $\afIPDS : \syn{Exp} \to \mathbb{IPDS}$}.
                  \label{fig:acesk-to-ipds}
           \end{minipage}
          \end{figure*}

            \subsection{Introspective reachability via Dyck state graphs}
              \label{sec:pdreachability}
          
             \label{sec:fixpoint-dsg}
             Compiling an introspective pushdown system into a Dyck state graph 
             for exception-flow analysis does not require special modification
             with repect of the iterative method:
             The function $\mkDSG : \mathbb{IPDS} \to (\mathbb{DSG} \to
                 \mathbb{DSG})$ generates the monotonic iteration function we need:
             \begin{align*}
             \mkDSG(M) &= f\text{, where }
             \\
                 M &= (\QStates,\StackAlpha,\transfunction,\qstate_0)
                 \\
                   f(\DSStates,\DSFrames,\DSEdges,\dsstate_0) &= (\DSStates',\DSFrames,\DSEdges',\dsstate_0) \text{, where }
                   \\
                     \DSStates' &= \DSStates \union \setbuild{ \dsstate' }{ 
                       \dsstate \in \DSStates 
                         \text{ and }  
                       \dsstate 
                         \mathrel{\RPDTrans_M}
                       \dsstate'
                     }
             \union \set{\dsstate_0}
             \\
                 \DSEdges' &= \DSEdges \union \setbuild{ \dsstate \pdedge^\stackact \dsstate' }{ 
                   \dsstate \in \DSStates 
                     \text{ and }  
                   \dsstate 
                     \mathrel{\RPDTrans_M^\stackact}
                   \dsstate'    
                 }
             \text.
             \end{align*}
             Our implementation of thefunction $\mathbb{DSG}$ correspondents exactly what's defined as above.
             In section \ref{subsec: dcg-exn}, we will show  details of computing Dyck state graph in the presence of exception flows.

\subsection{Allocation: Polyvariance, context-sensitivity and object-sensitivity}
\label{sec:polyvariance}
In the abstract semantics, 
the abstract allocation functions take the form: 
$\aallocFP: \syn{Stmt} \times \sa{Conf} \rightharpoonup \aFramePointer$
and $\aallocOP : \syn{Stmt} \times \sa{Conf}  \rightharpoonup \aObjectPointer $.
The two allocation functions determine the polyvariance and object-sensitivity of the analysis.
(In control-flow analysis, \textit{polyvariance} literally referes to 
the number of abstract addresses(variants) there are for each variable.)
All of the following allocation approaches can be used with abstract semantics:
\begin{itemize}
\item{\textit{\textbf{Monovariance: Pushdown 0CFA}}}
Pushdown 0CFA passes the statement itself for abstract addresses, 
meaning that $\aFramePointer$ will be passed the call site statement, 
and $\aObjectPointer$  the instantiation site statement:
\begin{align*}
\aFramePointer &= \syn{Stmt} & & & \aObjectPointer &= \syn{\syn{Stmt}} 
\\
\aallocFP(s, \aconf) &= s & & & \aallocOP(\mathit{s}, \aconf)&= \mathit{s} 
\end{align*} 
\item{\textit{\textbf{Pushdown 1CFA}}}
Pushdown 1CFA pairs the statement with current statement to get an abstract address:
\begin{align*}
\aFramePointer &= \syn{Stmt} \times \syn{Stmt}   
\\
\aallocFP(s, (\vec{s'}, \afp, \astore, \acont)) &= (s, s'_0) 
\\
\aObjectPointer &= \syn{Stmt} \times \syn{Stmt} 
\\
 \aallocOP(\mathit{s}, (\vec{s'}, \afp, \astore, \acont))&= (\mathit{s}, s'_0)
\end{align*} 

\item{\textit{\textbf{Pushdown $k$-CFA}}} 
Pushdown $k$-CFA looks beyond the current state and at the last $ k$ states. 
By concatenating the statements in the last $k$ states together, and pairing this sequence 
with a variable we get pushdown $k$-CFA:
\begin{align*}
\aFramePointer &= \syn{Stmt} \times \syn{Stmt}^k   
\\
\aallocFP(s, \vect{(\vec{s_1}, \afp, \astore, \acont), \dots}) &= (\mathit{s}, \vect{s_{10}, \dots s_{k0} })
\\
\aObjectPointer &= \syn{Stmt} \times \syn{Stmt}^k
\\
 \aallocOP(\mathit{s}, \vect{(\vec{s_1}, \afp, \astore, \acont), \dots})&= (\mathit{s}, \vect{s_{10}, \dots s_{k0} })
\end{align*} 
\end{itemize}
In addition, there is much static \textit{context} information after getting Abstract Syntax Tree(AST), 
such as for each statement, we can know its line number, what class and method it belongs to.
By default, we also take advantage and instrument these information  as complementary to the above \textit{context}
formalized.

\subsection{System architecture}

\begin{figure*}
\begin{center}
\includegraphics[scale = 0.9]{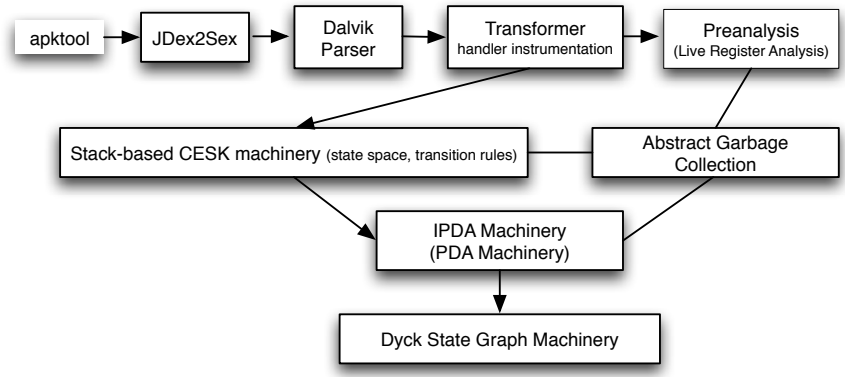}
\caption{System Architecture of (i)pushdown exception flow analysis.
         (Lines without arrows indicates the components 
          are implicitly connected)}
\label{fig:impl-arch}
\end{center}
\end{figure*}

We have implemented the analytic framework in Scala. Figure \ref{fig:impl-arch}
presents the system architecture:
{\sf{apktool}} extracts {\tt{.dex}} file from Android applications~\cite{local:apktool:url}.
{\sf{JDex2Sex}} extracts class files from the {\tt{.dex}} file  to generate an S-expression encoding 
the {\tt dex file}.
The S-expression IR is then fed into {\sf{Dalvik Parser}} and parsed into a Dalvik AST.
The {\sf{Transformer}} takes another pass on the Dalvik AST to instrument   
$\syn{push}\mhyphen\syn{handler}$ statements and $\syn{pop}\mhyphen\syn{handler}$ pseudo-statements,
and attach some other context information to statements.
{\sf{Preanalysis}}, specifically, live register analysis,  is performed right after {\sf{Transformer}}. 
It is an intra-procedural backward data flow analysis on instructions for each method~\cite{local:new-dragon}.

The core pushdown analytic components starts from the second row in Fig \ref{fig:impl-arch}.
The implementation of each component follows its correspondent formulation:
{\sf{Stack-based CESK machine}} embodies the abstract state space as shown in Fig~\ref{fig:abs-conf-space}, 
and abstraction transition relations in Section~\ref{sec:abs-transition-rules}.
{\sf{(I)PDCFA Machinery}} injects 
the program into a rooted pushdown system
(Figure~\ref{fig:acesk-to-ipds}).
A {\tt --gc} flag determines whether we use {\sf{PDCFA Machinery}} or {\sf{(I)PDCFA Machinery}}.
{\sf{Dyck State Graph Machinery}} implements
the fixed-point synthesis algorithm
(summarized in in Appendix~\ref{sec:fixpoint-dsg}).
In the following section, we will focus on the details of summarization algorithm in this machinery in handling exception flows.

\end{document}